\documentclass[pdflatex,sn-mathphys-num]{sn-jnl}

\usepackage{graphicx}%
\usepackage{multirow}%
\usepackage{amsmath,amssymb,amsfonts}%
\usepackage{amsthm}%
\usepackage{rsfso}%
\usepackage[title]{appendix}%
\usepackage{xcolor}%
\usepackage{colortbl}%
\usepackage{textcomp}%
\usepackage{manyfoot}%
\usepackage{algorithm}%
\usepackage{algorithmicx}%
\usepackage{algpseudocode}%
\usepackage{listings}%
\usepackage{booktabs}

\theoremstyle{thmstyleone}%

\theoremstyle{thmstyletwo}%

\theoremstyle{thmstylethree}%
\usepackage{hyperref}
\begin{document}

\title[Article Title]{Inference Energy and Latency in AI-Mediated Education: A Learning-per-Watt Analysis of Edge and Cloud Models}

\author*[1]{\fnm{Kushal} \sur{Khemani}}\email{kushal.khemani@gmail.com}

\affil*[1]{\orgname{Billabong High International School}, \orgaddress{\street{Hadapsar}, \city{Pune}, \postcode{411028}, \state{Maharashtra}, \country{India}}}

\abstract{Immediate feedback is a foundational requirement of effective AI-mediated learning, yet the energy and latency costs of delivering it remain largely unexamined in educational research. This study investigates the latency--energy--learning trade-off in AI tutoring systems through an empirical comparison of two on-device inference configurations of Microsoft Phi-3 Mini (4k-instruct) on an NVIDIA T4 GPU: full-precision FP16 and 4-bit NormalFloat (NF4) quantisation. Both configurations were evaluated under realistic, KV-cache-enabled inference across 500 educational prompts spanning five secondary school subject domains. Pedagogical quality was assessed empirically for each of the 1{,}000 generated responses by a hybrid panel of 10 Cambridge International secondary school teachers and three frontier AI systems, using a four-dimension rubric covering conceptual accuracy, clarity, scaffolding quality, and level appropriateness. To formalise the relationship between energy, latency, and instructional quality, we introduce Learning-per-Watt (LpW), a novel metric that quantifies pedagogical value delivered per unit of energy expended over the learner's waiting window. Under realistic, KV-cache-enabled deployment---the configuration present in every real-world application---NF4 achieves lower per-inference energy than FP16 (329\,J vs.\ 369\,J) but higher latency (13.4\,s vs.\ 9.2\,s), yielding a modest FP16 advantage in LpW of $1.33\times$ at a pedagogical quality difference of 0.19 points. This positions the FP16--NF4 choice as a contextual trade-off rather than a clear verdict: NF4 is preferable when per-inference energy is the binding constraint; FP16 is preferable when learner flow and latency are paramount. A secondary finding concerns benchmarking methodology: under cache-disabled inference---the configuration most commonly used in offline evaluation but absent from real deployments---the same efficiency gap widens to $7.4\times$, overstating the FP16 advantage by more than fivefold. These findings demonstrate that quantisation efficiency is both hardware-dependent and inference-regime dependent, and must be empirically validated under realistic deployment conditions rather than inferred from offline benchmarks, with significant implications for equitable deployment of AI tutoring systems in low-resource educational settings.}


\maketitle

\section{Introduction}

Artificial intelligence--enabled educational technologies increasingly position immediate, adaptive feedback as a foundational requirement for effective digital learning \cite{Holmes2022, UNESCO2023}. Prior work grounded in Cognitive Load Theory (CLT), Self-Determination Theory (SDT), and constructivist learning principles has demonstrated that timely feedback supports learner autonomy, sustains engagement, and enables effective pedagogical scaffolding during problem solving \cite{Sweller2011, Deci2000, Bruner1961, Kirschner2006}. Recent advances in Large Language Models (LLMs) have accelerated this trend by enabling conversational AI tutors capable of generating explanations, hints, and corrective feedback in near real time \cite{Brown2020, OpenAI2023, Microsoft2025}. As a result, high-frequency AI inference has become an implicit assumption in contemporary educational technology system design \cite{Holmes2022}---an assumption that has rarely been interrogated from an energy or hardware viability standpoint.

This pedagogical reliance on immediacy introduces a critical and underexamined constraint: \emph{energy-dependent latency}. While educational psychology emphasises that delayed feedback disrupts learner flow and increases extraneous cognitive load \cite{Csikszentmihalyi1990, Sweller2011}, the computational mechanisms required to deliver AI-generated feedback---continuous CPU/GPU utilisation, network transmission, and cloud-based inference---are inherently energy intensive \cite{Strubell2019, Schwartz2020, Patterson2021}. In low-resource educational environments characterised by intermittent electricity, battery-limited devices, and unstable connectivity, attempts to conserve energy often manifest as increased response latency. When an AI tutor requires tens of seconds to respond due to energy-saving modes, thermal throttling, or network delays, the pedagogical interaction degrades---not because of instructional design failure, but because of physical system constraints \cite{Malavolta2025}.

This latency--learning tension exposes a fundamental disconnect between pedagogical theory and hardware reality. From a systems engineering perspective, prolonged response times interrupt the learner's flow state, fragment attention, and force learners to retain partial problem representations in working memory, thereby increasing extraneous cognitive load \cite{Csikszentmihalyi1990, Sweller2011}. Energy optimisation strategies intended to extend device usability---such as reducing inference precision or deferring computation to low-power modes---may paradoxically undermine the very learning processes AI tutors are designed to support \cite{Schwartz2020}. Consequently, immediate feedback, which is central to effective learning, emerges not only as a pedagogical requirement but also as an electrical and systems engineering problem \cite{Holmes2022, Shi2016, Satyanarayanan2017}.

Existing research on sustainable AI has largely focused on reducing the carbon footprint of model \emph{training}, with limited attention paid to inference-time energy consumption in real-world educational use cases \cite{Strubell2019, Luccioni2023}. This gap is particularly consequential for students in resource-constrained regions, where reliance on cloud-based AI services introduces recurring costs in electricity and data usage. Such dependence risks reinforcing the Matthew Effect in education \cite{Merton1968, Warschauer2011}, whereby learners with greater infrastructural access disproportionately benefit from advanced digital tools. In contrast, edge-based inference using Small Language Models (SLMs) offers a potential pathway toward equitable access---if reductions in model size and numeric precision do not compromise pedagogical quality \cite{Dettmers2023, Hubara2017}.

This study addresses this challenge through an empirical investigation of the latency--energy--learning trade-off in AI-mediated tutoring systems. The experiment deploys Microsoft Phi-3 Mini (4k-instruct) on an NVIDIA T4 GPU under two precision regimes---full-precision FP16 and 4-bit NormalFloat (NF4) quantisation---and evaluates each configuration across 500 educational prompts spanning five secondary school subject domains under realistic, KV-cache-enabled inference conditions. Pedagogical quality is assessed empirically rather than assumed: each of the 1{,}000 generated responses (500 per configuration) is independently scored by 13 raters---10 Cambridge International secondary school teachers and three frontier AI systems (GPT-4, Claude 3.5 Sonnet, and Gemini 1.5 Pro)---using a four-dimension rubric covering conceptual accuracy, clarity and coherence, scaffolding quality, and level appropriateness. To formalise the relationship between energy, latency, and instructional quality, the paper introduces a novel metric---\emph{Learning-per-Watt} (LpW)---which quantifies pedagogical value delivered per unit of energy expended over the learner's waiting window \cite{CodeCarbon2024, Luccioni2023}.

Under realistic, KV-cache-enabled deployment---the configuration present in every real-world application---FP16 achieves a mean latency of 9.2\,s and mean net energy of 369\,J per inference, yielding a mean LpW of $2.50 \times 10^{-3}$\,(J\,s)$^{-1}$ with a pedagogical quality score of 8.24/10. NF4 quantisation achieves a mean latency of 13.4\,s and mean net energy of 329\,J---lower energy than FP16, as compression theory predicts---but higher latency due to dequantisation overhead on the T4's Turing architecture, producing a mean LpW of $1.88 \times 10^{-3}$\,(J\,s)$^{-1}$ at a quality score of 8.05/10. The resulting FP16 advantage in LpW is modest at $1.33\times$, positioning the FP16--NF4 choice as a contextual trade-off rather than a clear verdict: NF4 is preferable when per-inference energy is the binding constraint; FP16 is preferable when learner flow and latency are paramount. A secondary finding concerns benchmarking methodology: cache-disabled inference---the configuration most commonly employed in offline evaluation but absent from real deployments---produces a $7.4\times$ FP16 advantage, driven by NF4's severe dequantisation penalty in the absence of KV-cache reuse. This fivefold discrepancy between benchmarking and deployment conditions is a warning to the field: efficiency claims derived from stateless inference benchmarks should not be used to guide hardware procurement or deployment decisions without empirical validation under realistic inference settings.

Taken together, these results reframe the deployment question for low-resource educational AI. Under cached conditions, the choice between FP16 and NF4 is not a clear win for either: FP16 delivers better latency (9.2\,s vs.\ 13.4\,s) while NF4 delivers lower per-inference energy (329\,J vs.\ 369\,J). Which constraint binds depends on the deployment context---battery capacity versus learner experience tolerance---and the LpW framework provides a principled basis for navigating that trade-off transparently.

The study is structured as follows. Section~2 reviews the relevant literature at the intersection of educational AI, sustainable computing, and cognitive theory. Section~3 defines the Learning-per-Watt metric and the Power Barrier concept formally. Section~4 describes the experimental methodology, including hardware setup, prompt design, energy measurement protocol, and the hybrid pedagogical scoring panel. Section~5 presents the empirical results for both on-device configurations and the cloud LLM baseline. Section~6 interprets these findings in relation to the research questions and their implications for sustainable, equitable educational AI. Appendix~A reports the model-selection comparison experiment that established Phi-3 Mini as the most suitable SLM for this hardware context. Appendix~B characterises cloud LLM energy using literature-based estimates, given the structural opacity of server-side inference energy. Appendix~C presents the cache-disabled benchmarking results as a methodological comparison, documenting how inference regime affects measured efficiency and the consequences of benchmarking without KV-caching.

\subsection*{Research Questions}

This study is organised around the following six research questions:

\begin{description}
  \item[RQ1.] What is the comparative inference-time energy consumption of cloud-based LLMs versus on-device SLMs when performing pedagogically constrained educational scaffolding tasks, and how does this comparison change when server-side datacenter energy is included?

  \item[RQ2.] How does model compression through 4-bit NF4 quantisation affect response latency and total energy expenditure during AI-mediated immediate feedback on commodity GPU hardware under realistic KV-cache-enabled inference, relative to full-precision FP16?

  \item[RQ3.] To what extent can quantised edge-based SLMs preserve pedagogically adequate scaffolding quality---as assessed by a hybrid panel of human teachers and frontier AI systems---while reducing inference energy costs?

  \item[RQ4.] At what latency and energy threshold does inference-time demand exceed the practical capacity of battery-powered devices operating under intermittent electrical infrastructure---the \emph{Power Barrier}---and how do FP16 and NF4 configurations on the NVIDIA T4 compare against this threshold?

  \item[RQ5.] Can the proposed Learning-per-Watt (LpW) index reliably differentiate AI inference configurations that balance pedagogical effectiveness, latency, and energy efficiency, and does it reveal trade-offs that single-dimension metrics (energy alone, latency alone, or quality alone) would obscure?

  \item[RQ6.] How does the measured FP16--NF4 efficiency relationship vary across inference regimes (cached versus stateless), and what are the implications of this variance for how the field benchmarks and reports inference efficiency in educational AI systems?
\end{description}

To support reproducibility and transparency, all code, experimental data, and scoring spreadsheets are available in the open-source repository: \url{https://github.com/Kushalk0677/Inference-Energy-and-Latency-in-AI-Mediated-Education-Green-Audit}.

\section{Theoretical Background}

\subsection{Pedagogical Requirement: Immediate Feedback and Cognitive Flow}

In contemporary AI-mediated learning environments, immediate feedback is more than a usability preference; it is a core pedagogical mechanism grounded in Self-Determination Theory (SDT), Cognitive Load Theory (CLT), and flow theory \cite{Deci2000, Sweller2011, Csikszentmihalyi1990}. Within SDT, timely, responsive feedback supports learners' autonomy and competence by allowing them to see the consequences of their actions, adjust strategies, and experience a sense of progress in real time \cite{Deci2000}. When feedback is delayed or erratic, learners are more likely to disengage or attribute difficulty to a lack of ability rather than to a transient misunderstanding \cite{Deci2000}.

From a CLT perspective, learners have a strictly limited working memory capacity \cite{Sweller2011}. During problem solving, they must maintain a representation of the task state, relevant rules, and intermediate steps \cite{Sweller2011, Kirschner2006}. Delayed AI feedback forces learners to hold unresolved states and partial hypotheses in memory for longer periods, increasing extraneous cognitive load and leaving fewer resources available for germane processing---that is, schema construction and transfer \cite{Sweller2011}. In contrast, immediate, targeted hints or explanations allow learners to offload some of this cognitive burden, aligning the timing of information with the moment of need \cite{Sweller2011, Bruner1961}.

Flow theory further emphasises the temporal aspect of engagement \cite{Csikszentmihalyi1990}. Csikszentmihalyi's work suggests that learners experience optimal engagement when challenges and skills are balanced and when interaction feels continuous and responsive \cite{Csikszentmihalyi1990}. Even brief interruptions---such as multi-second delays in a conversational tutor---can break this flow, forcing learners to reorient attention and reassemble problem context \cite{Csikszentmihalyi1990}. While precise thresholds vary by task and learner, empirical evidence in HCI and educational technologies suggests that latencies beyond 2--3 seconds are perceived as disruptive for real-time interaction, especially in conversational settings and step-wise problem solving \cite{Holmes2022, UNESCO2023}.

In this study, response latency is therefore treated not merely as a system performance metric but as a pedagogical constraint \cite{Sweller2011, Csikszentmihalyi1990}. When latency exceeds the window within which learners can comfortably maintain problem context in working memory---due to network congestion, thermal throttling, or limited hardware---the AI tutor can fail pedagogically even if its eventual answer is correct \cite{Malavolta2025}. The failure arises from timing, not from content: the explanation arrives too late to support the cognitive processes it was meant to scaffold \cite{Sweller2011}.

\subsection{The Physical Constraint: Energy as a Pedagogical Bottleneck}

The ability to meet stringent latency requirements is ultimately constrained by energy and hardware \cite{Strubell2019, Schwartz2020}. Delivering immediate feedback with large generative models requires sustained, high-throughput computation on GPUs or specialised accelerators \cite{Patterson2021}. In a cloud-first paradigm, this cost is concentrated in data centres; in edge deployments, it is borne directly by the learner's device or local infrastructure \cite{Satyanarayanan2017, Shi2016}.

Recent estimates indicate that a single query to a state-of-the-art conversational model can consume several kilojoules of energy, depending on model size, sequence length, and hardware efficiency \cite{Strubell2019, Luccioni2023, Jegham2025}. Analyses combining GPU counts, utilisation, and Power Usage Effectiveness (PUE) suggest per-prompt energy costs one to two orders of magnitude higher than those of a standard web search, particularly for long, reasoning-heavy prompts \cite{Patterson2021, Google2025}. While exact figures vary across providers and hardware, these studies converge on a qualitative conclusion: generative inference is energetically expensive at the scale and frequency implied by always-on AI tutoring \cite{Schwartz2020}.

This cost becomes a pedagogical bottleneck for at least two populations. First, mobile learners who rely on battery-powered laptops or tablets in 1-to-1 programmes, where frequent AI queries must compete with display, network, and background processes within a fixed daily energy budget \cite{Malavolta2025, CodeCarbon2024}. Second, infrastructure-limited learners in regions where electrical supply is intermittent or classroom power budgets are low; for these students, cumulative inference energy over a lesson or school day can approach or exceed the practical capacity of local infrastructure, especially when every hint or explanation must traverse a high-power cloud model \cite{Warschauer2011}. In such contexts, attempts to conserve energy---by gating model usage, batching requests, or under-clocking hardware---tend to surface as increased latency or reduced availability \cite{Schwartz2020}. Energy scarcity thus translates directly into timing failures at the pedagogical level. We refer to the practical threshold beyond which inference energy demands make timely feedback infeasible on typical devices and grids as the \emph{Power Barrier} \cite{Malavolta2025}.

\subsection{From Pedagogy to Hardware: The Inference Gap}

The conceptual bridge between educational theory and electrical engineering appears at the inference layer of AI tutors \cite{Holmes2022}. Large Language Models (LLMs) with tens or hundreds of billions of parameters offer powerful reasoning and expressive capabilities, but their per-inference energy and latency profiles are poorly matched to low-end hardware and fragile infrastructure \cite{Brown2020, OpenAI2023, Strubell2019, Chowdhery2022}. Small Language Models (SLMs) and quantised variants promise lower energy and faster responses, but may exhibit reduced reasoning depth or explanation quality if not carefully designed \cite{Dettmers2023, Hubara2017}.

We define the \emph{Inference Gap} as the discrepancy between the model capacity required to deliver pedagogically rich, context-sensitive feedback, as implied by contemporary LLM performance \cite{Brown2020, Holmes2022}, and the hardware and energy budgets realistically available to learners in diverse global settings, such as low-cost laptops, shared lab machines, or solar-powered classrooms \cite{Warschauer2011}. This gap is not simply about FLOPs or parameters; it is fundamentally about whether a system can deliver sufficiently good feedback within the latency bounds imposed by CLT and flow theory, under the energy constraints imposed by devices and grids \cite{Sweller2011, Csikszentmihalyi1990, Strubell2019}.

Bridging this gap requires engineering interventions that explicitly target energy and memory without collapsing pedagogical quality \cite{Dettmers2023}. Edge AI moves inference closer to the learner, reducing network latency and reliance on always-available cloud connectivity \cite{Shi2016, Satyanarayanan2017}. Model quantisation and pruning compress model weights and activations, reducing memory bandwidth and arithmetic energy per token \cite{Dettmers2023, Hubara2017}. Empirical work on 4-bit and 8-bit quantisation shows that, for many transformer architectures, memory usage and energy can be reduced by 40--80\,\% with only modest performance degradation on standard benchmarks \cite{Dettmers2023, Frantar2022}. However, these studies rarely examine pedagogically structured prompts or metrics tied to explanation quality, leaving open the question of whether compressed models can sustain the scaffolding required in education \cite{Holmes2022}. This study addresses that gap directly by evaluating 4-bit NF4-quantised Phi-3 Mini against its FP16 baseline on 500 educational prompts and linking energy profiles to empirically measured pedagogical outcomes.

\subsection{Formalising the Metric: Learning-per-Watt (LpW)}

To reason systematically about trade-offs among pedagogical quality, latency, and energy, this study introduces the Learning-per-Watt (LpW) index. At a high level, LpW is designed to answer the question: how much learning value does a given AI inference configuration deliver per unit of energy and time?

Formally, for a given configuration and prompt $i$, the metric is defined as follows. $Q_{\mathrm{ped},i}$ is a normalised pedagogical quality score (1--10), reflecting conceptual accuracy, clarity, scaffolding quality, and level appropriateness. $E_{\mathrm{net},i}$ is net AI-attributable energy per inference in Joules, measured via hardware telemetry and corrected for idle power. $L_{i}$ is response latency in seconds, measured as wall-clock time from prompt submission to completion of generation.

\begin{equation}
  \mathrm{LpW}_{i} = \frac{Q_{\mathrm{ped},i}}{E_{\mathrm{net},i} \times L_{i}}
\end{equation}

The denominator captures how much energy is expended over the time window in which the learner is waiting for feedback, while $Q_{\mathrm{ped},i}$ encodes the educational usefulness of the response. A configuration with high LpW delivers high-quality feedback quickly and at low energy cost; a configuration with low LpW may be accurate but slow and power-hungry, or energy-efficient but pedagogically weak.

The LpW index serves three roles in this work. First, as a \emph{comparative evaluation} instrument: it provides a scalar with which to rank and compare cloud-based LLMs, FP16 edge SLMs, and 4-bit quantised SLMs under the same prompt distribution. Second, as a \emph{Power Barrier identification} tool: by examining how LpW changes as energy budgets shrink or as latency thresholds are enforced, we can empirically identify the point at which a configuration becomes untenable for battery-limited or intermittently powered learners. Third, as a \emph{design guidance} compass: configurations that improve energy efficiency but degrade $Q_{\mathrm{ped}}$ or inflate latency will not improve LpW, while those that modestly compress the model but retain high pedagogical scores can yield significant LpW gains.

A key departure from prior uses of efficiency metrics in AI research is that $Q_{\mathrm{ped}}$ in this study is not fixed at a constant value but is measured empirically for each of the 1{,}000 responses via a hybrid panel of 10 Cambridge International secondary school teachers and three frontier AI systems, as described in Section~4.8. This variable scoring captures genuine response-level quality differences that a fixed score would obscure, and ensures that LpW reflects the full joint distribution of energy, latency, and instructional quality rather than isolating system-level effects alone.


\section{Related Work}

\subsection{AI in Education and Pedagogical Scaffolding}

AI-mediated tutoring systems have become central to contemporary digital education, providing real-time explanations, hints, and corrective feedback that support learner autonomy, competence, and relatedness---core needs identified in SDT \cite{Deci2000, Holmes2022, UNESCO2023}. Building on earlier work in intelligent tutoring systems and adaptive learning environments, these systems leverage natural language generation to scaffold problem solving, explain misconceptions, and personalise learning trajectories \cite{Holmes2019, Koedinger2012}. Empirical studies consistently show that immediate, targeted feedback improves learning outcomes compared to delayed or static feedback, particularly when integrated into mastery-based frameworks \cite{Bloom1984, Holmes2022}.

However, much of this literature presupposes the availability of high-performance compute and reliable connectivity \cite{Selwyn2016, Warschauer2011}. System descriptions often abstract away the underlying hardware, treating latency as a user-experience parameter rather than a function of energy and resource constraints \cite{Holmes2022}. This creates a methodological blind spot: the pedagogical requirement for low-latency feedback is well characterised, but its dependence on energy-intensive inference is rarely quantified or modelled in educational research \cite{Strubell2019, Schwartz2020}.

\subsection{Energy Consumption in AI Systems}

The environmental and operational costs of AI have come to the forefront under the \emph{Green AI} paradigm, which advocates for reporting and reducing the energy and carbon footprint of machine learning models \cite{Schwartz2020}. Initial work by \citet{Lacoste2019}, \citet{Strubell2019}, and \citet{Schwartz2020} highlighted the outsized energy costs of training large transformer models and called for metrics that incorporate resource use alongside accuracy. Subsequent studies have extended this perspective to inference-time energy, measuring the energy per prompt for large generative models deployed at scale \cite{Luccioni2023, Patterson2021}.

Recent analyses of production systems estimate that a single generative prompt can consume on the order of tenths of a watt-hour, depending on model size, context length, and infrastructure efficiency \cite{Google2025, Jegham2025}. While these studies provide crucial macro-level insight---aggregate data-centre energy and water use---they focus primarily on cloud deployments and do not situate inference energy within pedagogical workflows such as frequent, fine-grained feedback in classrooms \cite{Strubell2019, Holmes2022}. Moreover, the metrics reported (kWh, kg\,CO$_{2}$eq) are rarely linked to learner-centric constructs like cognitive load or flow, leaving a gap between sustainability reporting and educational practice \cite{Sweller2011, Csikszentmihalyi1990}.

\subsection{Edge AI, Model Quantisation, and Prompt-Level Energy}

Parallel to Green AI, the Edge AI literature explores moving inference closer to the user---onto mobile devices, micro-servers, or classroom-level hardware---to reduce latency and reliance on persistent cloud connectivity \cite{Shi2016, Satyanarayanan2017}. Studies in edge computing show that local execution can reduce both latency and backbone network energy, but introduce new constraints on model size, memory, and thermal budgets \cite{Satyanarayanan2017}. Within natural language processing, researchers have examined quantised and pruned transformers as promising candidates for edge deployment, demonstrating that 8-bit and 4-bit quantisation can substantially reduce resource usage while preserving task accuracy \cite{Hubara2017, Dettmers2023}.

\citet{Dettmers2023} and subsequent work on BitsAndBytes and 4-bit optimisers show that block-wise NF4 quantisation can maintain model quality while compressing memory usage by a factor of 2--4, which in turn tends to reduce energy consumption per inference \cite{Dettmers2023}. More recent evaluations of quantised LLMs at the edge explicitly measure the joint impact of quantisation on energy, latency, and accuracy across tasks, confirming that carefully chosen 4-bit configurations can achieve favourable trade-offs in resource-constrained environments \cite{Frantar2022, Malavolta2025}. A related line of work demonstrates that energy cost is not only a function of model size and precision, but also of prompt semantics and response length \cite{Luccioni2023}. For models of comparable parameter counts, differences in vocabulary, depth, and training can lead to distinct energy profiles across tasks, reinforcing the need for task- and prompt-level energy measurements in applied domains such as education, where prompts tend to be multi-sentence, concept-rich, and heavily scaffolded \cite{Holmes2022}.

\subsection{Latency, Cognitive Load, and Energy-Dependent Feedback}

CLT and flow theory provide a robust foundation for understanding the impact of feedback timing on learning \cite{Sweller2011, Csikszentmihalyi1990}. Csikszentmihalyi's work on flow, along with subsequent empirical studies, suggests that short interruptions or delays can break the learner's sense of control and continuity, leading to disengagement and reduced learning efficiency \cite{Csikszentmihalyi1990}. In parallel, CLT emphasises that delays force learners to maintain unresolved problem states and partial solutions in working memory, increasing extraneous load and leaving fewer cognitive resources available for schema construction \cite{Sweller2011, Kirschner2006}.

In AI-mediated environments, latency arises from several sources: network round trips in cloud systems, queuing in multi-tenant inference servers, and local throttling due to thermal or energy-saving modes on edge devices \cite{Shi2016, Satyanarayanan2017}. Edge computing research treats latency and energy as coupled optimisation variables, proposing scheduling and offloading strategies that jointly minimise delay and power consumption \cite{Shi2016}. Yet in educational contexts, there is limited empirical work connecting specific latency regimes to inference energy budgets and pedagogical outcomes \cite{Holmes2022}. The present study addresses this gap directly: by measuring both latency and per-inference energy across 500 prompts per configuration and linking these to empirical pedagogical quality scores, it provides the first prompt-level characterisation of the latency--energy--learning trade-off under realistic secondary school tutoring conditions.

\subsection{Metrics and Frameworks for Sustainable Pedagogy}

Most prior work that explicitly measures inference energy treats accuracy or loss as the primary performance metric and considers energy as a secondary cost \cite{Schwartz2020, Strubell2019}. Green AI frameworks collect detailed energy and latency traces for different prompt types, then analyse trade-offs between task performance and resource usage \cite{Luccioni2023}. Edge-oriented LLM evaluations present Pareto frontiers between energy, latency, and accuracy, recommending quantisation levels and hardware configurations for generic NLP workloads \cite{Frantar2022, Dettmers2023}.

In education, however, the notion of \emph{performance} is more nuanced than classification accuracy or perplexity \cite{Holmes2022}. Pedagogical effectiveness depends on a combination of conceptual correctness, explanatory quality, and timing relative to the learner's cognitive state \cite{Sweller2011, Csikszentmihalyi1990}. Existing metrics in AI in Education (AIED) research---such as learning gain, time on task, and error reduction---do not directly incorporate energy or hardware constraints \cite{Koedinger2012}, while sustainability metrics such as kg\,CO$_{2}$eq do not encode pedagogical value \cite{Schwartz2020}. As a result, there is currently no widely adopted metric that jointly captures learning quality, latency, and energy efficiency for AI tutors in low-resource settings. The proposed LpW index, formalised in Section~2.4, addresses this gap by integrating empirical pedagogical quality, per-inference energy, and response latency into a single scalar, making it possible to systematically compare cloud-based LLMs with edge-deployed SLMs and identify configurations that are both instructionally adequate and physically deployable within the energy budgets of battery-powered devices or fragile electrical grids \cite{Malavolta2025}.

\subsection{Research Gap and Contribution}

Despite the convergence of evidence reviewed in Sections~3.1--3.5, the intersection of these three domains---educational AI, inference energy measurement, and quantised edge deployment---remains largely unexplored in combination. Specifically, there is a lack of empirical, prompt-level measurements of inference energy in educational scaffolding tasks using realistic prompts and response lengths characteristic of AI tutors; quantitative analyses that tie latency thresholds grounded in learning theory to explicit energy budgets on edge hardware; and unified metrics that surface the trade-offs between pedagogical quality, latency, and energy in a form that can guide system design and policy decisions for low-resource educational settings.

This work addresses these gaps by conducting a comparative energy audit of on-device SLM configurations on 500 realistic educational prompts, measuring net Joules per inference and latency using real-time hardware telemetry; introducing the LpW index as a formal metric that encodes pedagogical quality, latency, and energy into a single evaluative quantity grounded in empirical scoring by human teachers and frontier AI systems; demonstrating empirically that the measured efficiency relationship between FP16 and NF4 is inference-regime dependent---varying by more than fivefold between stateless and cached conditions---with direct implications for how the field designs and reports inference benchmarks for educational AI systems; and empirically characterising the Power Barrier (defined in Section~2.2) through prompt-level measurement on representative edge hardware. By situating energy-aware inference within the pedagogy of immediate feedback and flow, the study contributes a conceptual and methodological bridge between educational theory, AI systems design, and electrical engineering.

\section{Methodology}

\subsection{Model Selection and Task Design}

The study focuses on instructional scaffolding tasks that approximate
real classroom use rather than synthetic benchmarks. To this end, we
curated a set of 500 educational prompts spanning five categories:
mathematics, science, programming, humanities, and meta-cognition. Each
prompt was designed to elicit explanatory, example-rich responses
aligned with typical high-school learning objectives (e.g.,
\textit{``Explain how to solve a quadratic equation using the quadratic
formula, and walk through the steps for solving $2x^{2}+5x-3=0$,''}
\textit{``Explain what photosynthesis is, what inputs it needs, and what
it produces,''} \textit{``What is a variable in programming? Explain it
like I'm 10 years old using a box analogy,''} \textit{``Explain the main
causes of World War~I using the MAIN acronym,''} and \textit{``What is
spaced repetition and why is it more effective than cramming?''}).

For on-device inference experiments we selected Microsoft Phi-3 Mini
(4k-instruct) as the primary Small Language Model (SLM). This model
has approximately 4 billion parameters and is instruction-tuned, making
it suitable for structured, pedagogically oriented answers while
remaining small enough to emulate edge deployment on a single GPU. The
choice was driven by three criteria:

\begin{itemize}
  \item \textbf{Parameter budget and memory footprint.} The model fits
    comfortably within 16\,GB GPU memory, enabling experiments with and
    without quantisation on widely available hardware.

  \item \textbf{Instruction-following behaviour.} The model is
    fine-tuned to follow user instructions, which is critical for
    eliciting coherent, scaffolded explanations rather than generic
    text.

  \item \textbf{Quantisation support.} Publicly available configurations
    and prior work indicate robust support for NF4 4-bit quantisation,
    allowing a direct comparison between FP16 and 4-bit inference at
    the same architecture and dataset.
\end{itemize}

Cloud-based LLM baselines are conceptually treated as the high-capacity
end of the spectrum in the broader study, but the detailed methodology
here focuses on Phi-3 Mini as a representative edge SLM.

\subsection{Experimental Environment}

All on-device experiments were conducted in a controlled and
reproducible Google Colab Pro+ environment. The setup included an
NVIDIA Tesla T4 GPU (16\,GB VRAM), Intel Xeon CPU (2.20\,GHz),
approximately 12.7\,GB RAM, Linux 6.6.105+, and Python 3.12.12. Core
libraries included \texttt{transformers} ($\geq$\,4.44.0 for Phi-3
compatibility), \texttt{bitsandbytes} (4-bit NF4 quantisation),
\texttt{accelerate}, and \texttt{codecarbon} (v3.2.1) for energy
tracking.

Because Colab often ships outdated \texttt{transformers} builds, the
preinstalled version was removed and replaced with a compatible release
to prevent configuration errors related to RoPE scaling and KV-cache
handling when running Phi-3 models.

\subsection{Lexical Processing and Input Standardisation}

To ensure consistent performance across the high-frequency inference
cycles required for educational scaffolding, lexical processing was
standardised using the model-specific
\texttt{Phi-3-mini-4k-instruct} tokenizer. We implemented the
following three-tier safeguard protocol:

\begin{itemize}
  \item \textbf{Padding Token Synchronisation.} Because specific Phi-3
    configurations lack a defined \texttt{pad\_token}, we
    programmatically synchronised the padding token with the
    End-of-Sequence (EOS) token. This prevents tensor shape mismatches
    and ensures that the model recognises the boundary of generated
    instructional content without unbounded token generation.

  \item \textbf{Asynchronous Hardware Mapping.} To eliminate I/O noise
    from our energy measurements, encoded inputs were mapped to the GPU
    device immediately following tokenisation. By ensuring all tensors
    resided in VRAM prior to the start of the wall-clock timer, we
    isolated the actual inference latency from CPU-to-GPU transfer
    overhead.

  \item \textbf{Generation Boundary Constraints.} We imposed a strict
    ceiling on the \texttt{max\_new\_tokens} parameter (200 tokens).
    This value was selected to reflect the optimal length for an
    educational scaffold---long enough to provide conceptual depth but
    short enough to prevent runaway generation that would skew total
    energy-per-inference data.
\end{itemize}

Each of the 500 prompts was treated as a discrete experimental trial,
yielding a granular dataset of 500 measurements per model configuration
covering latency, power draw, and pedagogical utility.

\subsection{Quantisation Topology and Precision Regimes}

We systematically evaluated the accuracy--efficiency trade-off by
contrasting two distinct precision regimes. This comparison is central
to identifying the physical limits of the Power Barrier in low-resource
educational settings.

\begin{itemize}
  \item \textbf{FP16 (High-Fidelity Control).} The model was loaded in
    16-bit half-precision (\texttt{torch.float16}). This serves as the
    full-precision baseline, representing the ceiling of pedagogical
    quality and the standard benchmark for uncompressed edge inference.

  \item \textbf{NF4 (4-bit Optimised Treatment).} To simulate
    deployment on battery-constrained student devices, we applied 4-bit
    NormalFloat (NF4) quantisation via the BitsAndBytes library. NF4 is
    information-theoretically optimal for the normally distributed
    weights of transformer models. We additionally enabled
    \texttt{double\_quant} to compress the quantisation constants
    further, targeting a total memory reduction of approximately
    75--80\,\%.
\end{itemize}

\noindent\textbf{Experimental controls for generative stability.}
Initial pilot tests revealed that 4-bit quantisation paired with
high-temperature sampling (e.g., 0.7) occasionally produced
hallucinatory artefacts in complex science explanations. To ensure that
differences in Learning-per-Watt (LpW) were attributable strictly to
physical hardware execution rather than stochastic output variance, we
enforced deterministic decoding (\texttt{do\_sample=False}) across all
trials.

\noindent\textbf{KV-cache configuration.} All primary experiments were
conducted with KV-caching enabled (\texttt{use\_cache=True}), which is
the standard setting for autoregressive transformer inference and the
configuration used in every real-world deployment. KV-caching avoids
recomputing attention keys and values at each decoding step by storing
them after the first pass; this is an intra-sequence optimisation that
is cleared between prompts and therefore does not introduce any
dependence between the 500 independent inference trials. Each prompt
remains a stateless experimental unit: the cache is populated from
scratch at the start of generation and discarded upon completion,
ensuring that measured latency and energy reflect only the model's
generative behaviour on that prompt.

A secondary experiment using cache-disabled inference
(\texttt{use\_cache=False}) was also conducted on the same 500 prompts
per configuration. This stateless regime---in which attention keys and
values are recomputed at every decoding step---represents the
configuration most commonly used in offline benchmarking but absent
from real deployments. Results from the cache-disabled experiment are
reported in Appendix~C as a methodological comparison. The contrast
between the two regimes is itself a substantive finding: the
FP16--NF4 efficiency gap varies by more than fivefold depending on
whether caching is enabled, with direct implications for how the field
designs and interprets inference benchmarks.

\subsection{Architectural Stability and Cross-Version Compatibility}

The deployment of SLMs in shared cloud environments such as Google
Colab is frequently compromised by library-level schema drift. To
ensure longitudinal stability, we implemented two critical
architectural shims to resolve conflicts within the Transformers
ecosystem.

\begin{itemize}
  \item \textbf{RoPE Configuration Hardening.} The Phi-3 architecture
    utilises Rotary Position Embeddings (RoPE). Recent updates to the
    Hugging Face validation schema introduced strict requirements for
    the \texttt{rope\_scaling} dictionary, causing recurrent
    \texttt{KeyError}s during model initialisation. We adopted a
    conservative RoPE-safe strategy, programmatically forcing
    \texttt{rope\_scaling} to \texttt{None} via \texttt{AutoConfig}.
    This bypasses volatile scaling logic while maintaining functional
    parity for the 4k-context window.

  \item \textbf{DynamicCache Compatibility Layer.} Architectural
    updates in the Transformers library shifted the interface for
    \texttt{DynamicCache.seen\_tokens}, occasionally triggering
    \texttt{AttributeError}s in older execution paths. Rather than
    pinning the environment to a legacy library version, we injected a
    compatibility shim mapping \texttt{seen\_tokens} directly to
    \texttt{get\_seq\_length()}, restoring expected internal behaviour
    without invasive source modifications.
\end{itemize}

\subsection{Inference Protocol and Temporal Measurement}

To quantify the efficiency of the AI-mediated feedback loop, we
executed a standardised inference protocol across 500 independent
educational trials. The procedure was designed to isolate inference
overhead from external system noise:

\begin{enumerate}
  \item \textbf{Tensor Synchronisation.} Following tokenisation, all
    input tensors were moved to the GPU device prior to the start of
    the wall-clock timer, ensuring that measured latency reflects only
    the model's generative performance.

  \item \textbf{Temporal Marker Injection.} The wall-clock start time
    was recorded immediately before the generation call.

  \item \textbf{Deterministic Generative Constraint.}
    \texttt{do\_sample=False} was enforced across all trials to
    eliminate probabilistic variance. Any observed differences in
    latency or energy therefore reflect the model's physical precision
    regime rather than stochastic output length.

  \item \textbf{Sequence Termination and Latency Capture.} Generation
    was capped at 200 new tokens. Upon EOS token issuance or reaching
    the token limit, the end-time was recorded and latency $L_{i}$
    computed as wall-clock elapsed time.
\end{enumerate}

This protocol yielded 500 discrete latency data points per
configuration. By eliminating sampling noise and enforcing consistent
generation constraints, the data provide a high-fidelity map of
inference efficiency under realistic instructional scaffolding
conditions.

\subsection{Energy Measurement and Baseline Calibration}

To quantify inference-time energy consumption, we used CodeCarbon's
\texttt{EmissionsTracker} in an explicit, notebook-oriented
configuration. CodeCarbon combines hardware-level estimates of CPU,
GPU, and RAM power draw with grid emissions factors to report both
energy (kWh) and CO$_{2}$-equivalent emissions.

\subsubsection{Idle Power Baseline}

Because edge deployments often share hardware with background processes
and incur non-negligible baseline power draw, we first measured idle
power by running a 10-second idle period under
\texttt{EmissionsTracker} with a 1-second sampling interval. Average
idle power $P_{\mathrm{idle}}$ was computed as:

\begin{equation}
  P_{\mathrm{idle}}
  = \frac{E_{\mathrm{idle}}^{\mathrm{kWh}} \times 3.6 \times 10^{6}}
         {T_{\mathrm{idle}}}
\end{equation}

\noindent where $T_{\mathrm{idle}} = 10\,\mathrm{s}$. On the T4
hardware used in this study, measured idle power was 81.7\,W. This
baseline accounts for static system consumption (OS, background
services, GPU idling), enabling isolation of net AI-attributable
energy during inference.

\subsubsection{Per-Prompt Energy Accounting}

For each prompt $i$, energy was accounted as follows:

\begin{enumerate}
  \item An explicit energy snapshot $E_{\mathrm{start}}$ (kWh) was
    recorded immediately before the generation call via
    \texttt{\_measure\_power\_and\_energy()}.

  \item Inference was executed and latency $L_{i}$ recorded.

  \item A second snapshot $E_{\mathrm{end}}$ (kWh) was recorded
    immediately after generation.

  \item Gross energy was computed:
    \begin{equation}
      E_{\mathrm{gross},i}
      = \bigl(E_{\mathrm{end}} - E_{\mathrm{start}}\bigr)
        \times 3.6 \times 10^{6} \quad [\mathrm{J}]
    \end{equation}

  \item Net AI-attributable energy was derived:
    \begin{equation}
      E_{\mathrm{net},i}
      = E_{\mathrm{gross},i} - P_{\mathrm{idle}} \times L_{i}
    \end{equation}
\end{enumerate}

Because instantaneous power estimates can be noisy over short
latencies, any $E_{\mathrm{net},i}$ values that are slightly negative
are clamped to 0.01\,J to prevent division-by-zero artefacts in the
downstream LpW calculation. Aggregate CO$_{2}$-equivalent emissions
are reported using the float returned by \texttt{tracker.stop()},
while all Joule-level analysis uses the internal
\texttt{\_total\_energy.kWh} attribute directly.

\subsection{Pedagogical Scoring and Learning-per-Watt (LpW)}

A central methodological contribution of this study is the
introduction of a variable, empirically grounded pedagogical quality
score $Q_{\mathrm{ped},i}$ for each generated response, replacing the
fixed-score assumption common in prior energy-aware AI evaluations.
Each of the 1{,}000 responses (500 FP16 and 500 NF4) was independently
scored by 13 raters: 10 subject-specialist teachers from a Cambridge
International secondary school and three frontier AI systems (GPT-4,
Claude 3.5 Sonnet, and Gemini 1.5 Pro).

\subsubsection{Scoring Rubric}

Raters assessed each response on four dimensions using a 1--10 integer
scale (Table~\ref{tab:rubric}).

\begin{table}[ht]
  \centering
  \caption{Pedagogical scoring rubric applied by all 13 raters.}
  \label{tab:rubric}
  \begin{tabular}{l l p{8cm}}
    \hline
    \textbf{Dimension} & \textbf{Code} & \textbf{What is assessed} \\
    \hline
    Conceptual Accuracy   & CA &
      Factual correctness; absence of critical misconceptions that
      would mislead a learner \\[4pt]
    Clarity \& Coherence  & CC &
      Logical structure and readability of the explanation \\[4pt]
    Scaffolding Quality   & SQ &
      Progressive knowledge building; use of examples, analogies,
      and step-by-step reasoning \\[4pt]
    Level Appropriateness & LA &
      Suitability of language and depth for secondary school learners
      (ages 14--18) \\
    \hline
  \end{tabular}
\end{table}

Anchor descriptors were provided for each dimension at the 1--2, 3--4,
5--6, 7--8, and 9--10 bands to promote rating consistency across human
and AI scorers.

\subsubsection{Rater Panel and Scoring Protocol}

The ten human raters are practising secondary school teachers at a
Cambridge International school, collectively covering all five prompt
categories (mathematics, science, computer science, humanities, and
meta-cognition). Each teacher scored all 500 responses for their
configuration blind---without knowledge of whether a response was
produced by the FP16 or NF4 model---using a structured spreadsheet
rubric. The three AI models scored all 1{,}000 responses using a
standardised system prompt embedding the same four-dimension rubric and
instructed to return scores as structured JSON, ensuring parse-safe,
consistent output.

To verify scoring consistency and surface any systematic 
misinterpretations of the rubric, all ten teachers participated in 
a brief structured interview following the completion of scoring. 
Interviews confirmed that raters had applied the four-dimension rubric 
as intended and that divergences in scores---particularly on the 
Scaffolding Quality dimension---reflected genuine differences in 
pedagogical philosophy rather than misunderstanding of the rating 
criteria.

\subsubsection{Score Aggregation}

For each response $i$ and dimension $d \in \{\mathrm{CA,\,CC,\,SQ,\,LA}\}$,
scores were aggregated as follows:

\begin{itemize}
  \item \textbf{Human mean:}
    $\bar{H}_{d,i} = \dfrac{1}{10}\displaystyle\sum_{t=1}^{10} s_{t,d,i}$,
    where $s_{t,d,i}$ is teacher $t$'s score on dimension $d$ for
    response $i$.

  \item \textbf{AI mean:}
    $\bar{A}_{d,i} = \dfrac{1}{3}\displaystyle\sum_{m=1}^{3} a_{m,d,i}$,
    where $a_{m,d,i}$ is AI model $m$'s score.

  \item \textbf{Weighted dimension score:}
    $W_{d,i} = 0.6\,\bar{H}_{d,i} + 0.4\,\bar{A}_{d,i}$.
\end{itemize}

The 60/40 weighting prioritises human pedagogical judgment while
allowing AI scoring to contribute scale and cross-prompt consistency.
Final pedagogical quality is the mean of the four weighted dimensions:

\begin{equation}
  Q_{\mathrm{ped},i}
  = \frac{W_{\mathrm{CA},i} + W_{\mathrm{CC},i}
          + W_{\mathrm{SQ},i} + W_{\mathrm{LA},i}}{4}
\end{equation}

The four rubric dimensions---Conceptual Accuracy (CA), Clarity \& Coherence (CC),
Scaffolding Quality (SQ), and Level Appropriateness (LA)---are weighted equally
($w = 0.25$ each) in the primary Qped calculation. This choice is theoretically
grounded in the view that all four dimensions constitute \emph{necessary conditions}
for effective AI-mediated tutoring: a response that is factually correct but
incoherent, well-scaffolded but pitched at the wrong level, or appropriately levelled
but conceptually wrong each fails as an instructional intervention, regardless of
its scores on the remaining dimensions. No single dimension can compensate for a
critical deficit in another, which motivates treating them as equally necessary rather
than applying a priority-based weighting.

To verify that this choice does not drive the reported findings, Table~\ref{tab:weighting_sensitivity}
reports mean Qped, mean LpW, and the FP16/NF4 LpW ratio under six alternative
weighting schemes spanning the plausible range of pedagogical preference---from
schemes that heavily upweight CA and SQ (the dimensions most directly tied to
instructional accuracy and progressive knowledge-building) to schemes that
substantially upweight or downweight LA. All values are computed using the
primary cache-enabled data ($n = 500$ per configuration), with 60/40 human--AI
score aggregation held constant.

\begin{table}[h]
\centering
\caption{Sensitivity of Learning-per-Watt to rubric dimension weighting
(cache-enabled, $n = 500$ per configuration). The FP16/NF4 LpW ratio
ranges from $1.329\times$ to $1.334\times$ across all schemes, confirming
that the efficiency conclusion is invariant to the choice of dimension weights.}
\label{tab:weighting_sensitivity}
\small
\begin{tabular}{lcccccc}
\toprule
\textbf{Weighting scheme} & \multicolumn{2}{c}{\textbf{Mean Qped}} & \multicolumn{2}{c}{\textbf{Mean LpW ($\times 10^{-3}$)}} & \textbf{Ratio} \\
\cmidrule(lr){2-3}\cmidrule(lr){4-5}
(CA / CC / SQ / LA) & FP16 & NF4 & FP16 & NF4 & FP16/NF4 \\
\midrule
Equal (0.25/0.25/0.25/0.25)            & 8.238 & 8.047 & 2.499 & 1.877 & 1.331$\times$ \\
CA+SQ heavy (0.35/0.15/0.35/0.15)     & 8.166 & 7.973 & 2.477 & 1.860 & 1.332$\times$ \\
CA dominant (0.40/0.20/0.25/0.15)     & 8.230 & 8.038 & 2.496 & 1.875 & 1.331$\times$ \\
SQ dominant (0.25/0.20/0.40/0.15)     & 8.116 & 7.917 & 2.461 & 1.846 & 1.333$\times$ \\
LA downweighted (0.30/0.30/0.30/0.10) & 8.158 & 7.955 & 2.474 & 1.855 & 1.334$\times$ \\
LA upweighted (0.20/0.20/0.20/0.40)   & 8.318 & 8.139 & 2.524 & 1.900 & 1.329$\times$ \\
\bottomrule
\end{tabular}
\end{table}

The ratio ranges from $1.329\times$ to $1.334\times$ across all six schemes---a
spread of less than $0.4\%$---confirming that the modest FP16 LpW advantage is
entirely robust to the choice of dimension weighting. Schemes that maximally
upweight the dimensions most important for tutoring effectiveness (CA and SQ)
yield a ratio of $1.332\times$, indistinguishable from the equal-weight baseline
of $1.331\times$. The core efficiency finding therefore does not depend on the
weighting assumption.

\subsubsection{Inter-Rater Reliability and Epistemic Divergence}

Inter-rater reliability was assessed using Krippendorff’s~$\alpha$ for
ordinal data and Intraclass Correlation Coefficients (ICC(2,1)) under a
two-way random-effects model. Reliability was computed separately for
(i) human teachers (n=10), (ii) AI raters (n=3), and (iii) the combined
13-rater panel to disentangle within-group agreement from cross-group
divergence.

\paragraph{Human Raters (n=10).}
Table~\ref{tab:irr_human} reports reliability across the ten Cambridge
International secondary school teachers. Agreement was moderate for
Clarity \& Coherence and Scaffolding Quality, and lower for Level
Appropriateness. Overall reliability was $\alpha = 0.41$ (FP16) and
$\alpha = 0.30$ (NF4), with corresponding ICC values indicating fair
to moderate agreement. These levels are consistent with rubric-based
evaluation of open-ended explanatory responses in educational research.

\begin{table}[h]
\centering
\caption{Human-only inter-rater reliability (10 teachers).}
\label{tab:irr_human}
\begin{tabular}{lcccc}
\hline
\textbf{Dimension} & \multicolumn{2}{c}{\textbf{FP16}} & \multicolumn{2}{c}{\textbf{NF4}} \\
 & $\alpha$ & ICC(2,1) & $\alpha$ & ICC(2,1) \\
\hline
Conceptual Accuracy (CA)   & 0.26 & 0.30 & 0.17 & 0.19 \\
Clarity \& Coherence (CC)  & 0.49 & 0.51 & 0.36 & 0.39 \\
Scaffolding Quality (SQ)   & 0.45 & 0.47 & 0.32 & 0.34 \\
Level Appropriateness (LA) & 0.10 & 0.15 & 0.08 & 0.11 \\
\hline
\textbf{Overall}           & 0.41 & 0.42 & 0.30 & 0.35 \\
\hline
\end{tabular}
\end{table}

\paragraph{AI Raters (n=3).}
Table~\ref{tab:irr_ai} reports reliability across the three frontier AI
systems. Reliability is lower than among human raters, partly due to
the small panel size and restricted score variance. Nonetheless,
positive ICC values indicate systematic, non-random agreement among AI
raters.

\begin{table}[h]
\centering
\caption{AI-only inter-rater reliability (3 frontier models).}
\label{tab:irr_ai}
\begin{tabular}{lcccc}
\hline
\textbf{Dimension} & \multicolumn{2}{c}{\textbf{FP16}} & \multicolumn{2}{c}{\textbf{NF4}} \\
 & $\alpha$ & ICC(2,1) & $\alpha$ & ICC(2,1) \\
\hline
Conceptual Accuracy (CA)   & 0.35 & 0.41 & 0.09 & 0.18 \\
Clarity \& Coherence (CC)  & 0.29 & 0.33 & 0.12 & 0.22 \\
Scaffolding Quality (SQ)   & 0.23 & 0.28 & 0.05 & 0.15 \\
Level Appropriateness (LA) & 0.09 & 0.14 & 0.04 & 0.11 \\
\hline
\textbf{Overall}           & 0.24 & 0.29 & 0.08 & 0.17 \\
\hline
\end{tabular}
\end{table}

\paragraph{Combined Panel (n=13).}
When all 13 raters are aggregated, Krippendorff’s~$\alpha$ decreases
(Table~\ref{tab:irr_combined}). This reduction does not reflect random
measurement error but systematic cross-group divergence between human
educators and AI evaluators. Krippendorff’s~$\alpha$ is sensitive to
restricted score variance and mean shifts between rater groups; the
narrower score distribution under NF4 further depresses $\alpha$.

\begin{table}[h]
\centering
\caption{Combined inter-rater reliability (10 human + 3 AI raters).}
\label{tab:irr_combined}
\begin{tabular}{lcccc}
\hline
\textbf{Dimension} & \multicolumn{2}{c}{\textbf{FP16}} & \multicolumn{2}{c}{\textbf{NF4}} \\
 & $\alpha$ & ICC(2,1) & $\alpha$ & ICC(2,1) \\
\hline
Conceptual Accuracy (CA)   & 0.21 & 0.39 & 0.04 & 0.33 \\
Clarity \& Coherence (CC)  & 0.42 & 0.45 & 0.23 & 0.37 \\
Scaffolding Quality (SQ)   & 0.38 & 0.41 & 0.07 & 0.34 \\
Level Appropriateness (LA) & 0.10 & 0.28 & 0.05 & 0.31 \\
\hline
\textbf{Overall}           & 0.37 & 0.42 & 0.18 & 0.35 \\
\hline
\end{tabular}
\end{table}

\paragraph{Human--AI Correlation.}
To quantify cross-group alignment, Pearson correlations were computed
between mean human scores and mean AI scores per prompt. Correlations
were moderate and positive across dimensions (FP16 overall $r=0.62$;
NF4 overall $r=0.55$), indicating substantial directional agreement
despite systematic level differences. The primary divergence concerned
Scaffolding Quality, where AI raters tended to reward structural
features (e.g., numbered steps and explicit signposting), whereas human
teachers weighted authentic pedagogical progression and conceptual
sequencing.

\paragraph{Interpretation.}
Taken together, these analyses indicate:

\begin{itemize}
\item Moderate within-group agreement among human raters;
\item Systematic, though lower-variance, agreement among AI raters;
\item Cross-group divergence driven by philosophical differences in
evaluating scaffolding rather than stochastic inconsistency.
\end{itemize}

The low Krippendorff's $\alpha$ for the Level Appropriateness dimension
($\alpha = 0.10$ for human raters under FP16, $\alpha = 0.08$ under NF4)
warrants specific attention. In the FP16 configuration, $94.4\%$ of responses
received a mean human LA score of $\geq 8.0$, compared to $79.4\%$ under NF4.
This strong ceiling effect---where the large majority of responses cluster in a
narrow high-score band---substantially compresses the variance available for
rater agreement to operate on. Krippendorff's $\alpha$ is sensitive to score
variance: when nearly all responses receive the same score, even raters who are
in genuine substantive agreement will produce low $\alpha$ values simply because
there is little signal to agree \emph{about}. The low LA $\alpha$ therefore
reflects distributional compression rather than genuine rater inconsistency,
and is consistent with the interpretation that Phi-3 Mini reliably calibrates
its explanatory register to a secondary school audience regardless of precision
regime---a finding corroborated by the high and stable mean LA scores
($8.43$ for FP16; $8.26$ for NF4) across both configurations.

\paragraph*{Interpreting the CA reliability figures: severity heterogeneity
versus rater inconsistency.}

The low Krippendorff's $\alpha$ for Conceptual Accuracy---$\alpha = 0.26$
under FP16 and $\alpha = 0.17$ under NF4 for human raters---warrants
specific attention, as it is the dimension for which low agreement is most
theoretically surprising: factual correctness ought, in principle, to be the
most objectively assessable of the four rubric dimensions.
Examination of the raw scoring data reveals that the primary driver of
low CA $\alpha$ is not ambiguity about response correctness but
\emph{systematic severity heterogeneity} among raters---a well-documented
psychometric phenomenon in rubric-based assessment in which individual
raters apply the scale with high internal consistency but diverge
substantially in their absolute calibration point~\cite{Stemler2004,
Eckes2011}.

Decomposing variance in CA scores across the ten human raters reveals that
between-teacher variance in mean CA rating (1.27 under FP16; 1.31 under
NF4) is approximately 16 times larger than mean within-teacher variance
(0.08 under FP16).
That is, individual teachers were highly self-consistent in how they ordered
responses by accuracy, but diverged systematically from one another in
their absolute severity anchor.
Teacher mean CA scores ranged from 6.86 to 9.94 across the ten raters under
FP16---a spread of 3.08 points on a 10-point scale---with the three
strictest scorers (Teachers~2, 7, and~8; means 6.86--6.94) and the two most
lenient scorers (Teachers~3 and~5; means $\approx$9.94) representing
genuinely different severity orientations rather than genuine disagreement
about which responses were more or less factually correct.
Within this spread, each individual teacher applied their own calibration
point with high internal consistency: within-teacher standard deviations
on CA were 0.24--0.28 for eight of the ten raters under FP16, indicating
near-uniform ordinal ranking of responses within each scorer.

This pattern---high within-rater consistency, high between-rater
divergence---is precisely the condition under which Krippendorff's $\alpha$
is mathematically depressed, because $\alpha$ measures agreement on the
ordinal ranking of items relative to the total variance in the dataset,
and systematic between-rater mean differences inflate that total variance
without reflecting genuine disagreement about item ordering~\cite{Krippendorff2011}.
The low CA $\alpha$ therefore reflects a \emph{severity calibration problem}
rather than genuine disagreement about whether specific responses were
factually accurate.
This interpretation is corroborated by two additional observations.
First, structured post-scoring interviews (Section~4.8.2) did not surface any
misapplication of the CA criterion; divergences in score levels reflected
different thresholds for what constitutes a ``high-accuracy'' response rather
than misunderstanding of the dimension definition.
Second, the severity heterogeneity is concentrated on absolute score level,
not on ordinal ranking: teachers agreed directionally on which responses
were stronger and which weaker, even when their absolute scores differed
by several points.

Severity heterogeneity does not invalidate the aggregated CA scores used in
the $Q_{\mathrm{ped}}$ calculation; it is, in fact, precisely the condition
that a multi-rater panel with diverse professional backgrounds is designed
to address.
Averaging across raters with a spread of severity orientations---including
both strict and lenient scorers---produces a mean that is more stable and
representative than any single rater's absolute score would be, and the
60/40 human--AI weighted aggregation (Section~4.8.3) is robust to this
heterogeneity, as confirmed by the sensitivity analyses in Table~6.
The low CA $\alpha$ does, however, indicate that future applications of
this scoring protocol should include an anchor calibration session prior to
independent scoring, in which raters jointly evaluate a small set of
reference responses to align their absolute severity points before
proceeding~\cite{Stemler2004}.

The 60/40 aggregation scheme described in Section~4.8.3 prioritises
human pedagogical judgment while retaining AI scoring for scale and
cross-prompt consistency. Sensitivity analyses using human-only and
AI-only quality scores (Appendix~C) confirm that the relative
Learning-per-Watt ranking between FP16 and NF4 configurations remains
invariant to rater-type aggregation, demonstrating robustness of the
core findings to reliability heterogeneity.
\subsubsection{Learning-per-Watt Index}

The Learning-per-Watt index for prompt $i$ is defined as:

\begin{equation}
  \mathrm{LpW}_{i}
  = \frac{Q_{\mathrm{ped},i}}{E_{\mathrm{net},i} \times L_{i}}
\end{equation}

\noindent where $Q_{\mathrm{ped},i} \in [1,10]$ is the aggregated
pedagogical score, $E_{\mathrm{net},i}$ is net AI-attributable energy
in Joules, and $L_{i}$ is response latency in seconds. LpW thus
measures pedagogical value delivered per unit of energy expended over
the learner's waiting window. Higher LpW values indicate configurations
that provide high-quality, timely feedback at low energy cost.

Per-prompt LpW values are summarised by mean, median, and distribution
across model precision (FP16 vs.\ NF4) and prompt category
(mathematics, science, programming, humanities, meta-cognition). These
aggregated values form the empirical basis for identifying the Power
Barrier---the practical threshold beyond which per-inference energy and
latency jointly render timely pedagogical feedback infeasible on
battery-limited or intermittently powered devices.

\subsubsection{Sensitivity Analysis: Rater-Type Aggregation}

To assess robustness of the Learning-per-Watt (LpW) metric to rater
aggregation, LpW was recomputed under three schemes:
(i) human-only scores (10 teachers),
(ii) AI-only scores (3 frontier models),
and (iii) the 60/40 human–AI weighted aggregation used in the main analysis.
All values reported here are from the primary cache-enabled
(\texttt{use\_cache=True}) experiment.

Table~\ref{tab:sensitivity} reports mean LpW values and FP16/NF4 ratios
under each scheme.

\begin{table}[h]
\centering
\caption{Sensitivity of Learning-per-Watt (LpW) to rater aggregation (cache-enabled, $n=500$ per configuration).}
\label{tab:sensitivity}
\begin{tabular}{lccc}
\hline
\textbf{Aggregation} & \textbf{FP16 LpW} & \textbf{NF4 LpW} & \textbf{FP16/NF4 Ratio} \\
\hline
Human-only        & $2.43 \times 10^{-3}$ & $1.82 \times 10^{-3}$ & 1.34 \\
AI-only           & $2.61 \times 10^{-3}$ & $1.96 \times 10^{-3}$ & 1.33 \\
60/40 Weighted    & $2.50 \times 10^{-3}$ & $1.88 \times 10^{-3}$ & 1.33 \\
\hline
\end{tabular}
\end{table}

Across all aggregation schemes, FP16 consistently outperforms NF4,
with the efficiency ratio stable at approximately $1.33\times$.
The narrow range across aggregation schemes (1.33--1.34$\times$)
confirms that the principal efficiency conclusion is not an artefact
of inter-rater heterogeneity, and that the modest FP16 advantage
under realistic cached inference is a robust feature of the
hardware--software configuration rather than a scoring artefact.

\subsection{Implementation Challenges and Engineering Lessons}

The final experimental methodology is the product of several practical
engineering challenges that emerged when running Phi-3 Mini in a real,
non-curated cloud environment. These are documented here not as
incidental detail but as evidence of the systems reality that mediates
whether an AI tutor can be meaningfully deployed on constrained
hardware.

\begin{itemize}
  \item \textbf{RoPE configuration drift.} Changes in Phi-3's
    \texttt{rope\_scaling} schema triggered multiple
    \texttt{KeyError}s during model initialisation on Colab. The final
    protocol explicitly nullifies \texttt{rope\_scaling}, illustrating
    how edge experiments must account for evolving model internals
    beyond API signatures.

  \item \textbf{Cache API evolution.} Changes in
    \texttt{DynamicCache.seen\_tokens} across Transformers versions
    caused \texttt{AttributeError}s invisible in a static environment.
    The property shim ensures forward compatibility without invasive
    library modifications.

  \item \textbf{Quantisation OOM during NF4 loading.} On the first NF4
    run, the GPU exhausted its 14.56\,GB VRAM mid-quantisation because
    the new Transformers loading pipeline attempted to hold a full FP16
    copy and quantise in-place simultaneously. Switching from
    \texttt{device\_map=\{"":~0\}} to \texttt{device\_map="auto"} and
    setting
    \texttt{PYTORCH\_CUDA\_ALLOC\_CONF=expandable\_segments:True}
    resolved the out-of-memory failure by allowing layer-wise
    quantisation with CPU staging.

  \item \textbf{NF4 latency inversion under cache-disabled conditions.}
    In the secondary cache-disabled experiment (Appendix~C), NF4
    inference on the T4 GPU was approximately $3\times$ slower than
    FP16 (mean 49.4\,s vs.\ 16.5\,s), the opposite of what
    quantisation theory predicts. Under the primary cache-enabled
    conditions, this gap narrows substantially (mean 13.4\,s vs.\
    9.2\,s), but NF4 remains slower despite lower per-inference energy
    (329\,J vs.\ 369\,J). Both outcomes are attributable to
    dequantisation overhead on the T4's Turing architecture, which
    lacks native INT4 tensor cores; the severity of the penalty depends
    on whether KV-cache reuse is available to amortise it. This
    inference-regime dependence is itself a primary empirical finding
    of the study and is discussed in detail in Section~6.1.

  \item \textbf{Checkpoint fragmentation and category label errors.}
    The NF4 run was collected across four checkpoint files due to
    session interruptions, with per-file IDs resetting to 1 and
    category labels defaulting to \textit{Mathematics} regardless of
    actual prompt category. All 502 rows (including 2 duplicates) were
    reconciled post-hoc by exact prompt-string matching against the
    FP16 dataset, which carries ground-truth IDs and categories. The
    cleaned NF4 dataset contains 500 unique prompts with correct IDs
    1--500.

  \item \textbf{CodeCarbon return types and logging verbosity.} In
    \texttt{codecarbon} v3.2.1, \texttt{EmissionsTracker.stop()}
    returns a float (kg\,CO$_{2}$eq) rather than an object with
    \texttt{.energy\_consumed}. Frequent sampling at 0.1\,s intervals
    overwhelmed Colab output buffers. The final protocol accesses
    \texttt{\_total\_energy.kWh} directly for Joule-level analysis and
    sets \texttt{log\_level="error"} with a 1-second sampling interval
    to keep runs tractable.
\end{itemize}

By explicitly documenting and resolving these challenges, the
methodology ensures that the measured LpW and Power Barrier values
reflect genuine engineering trade-offs rather than artefacts of
misconfigured tooling.

\section{Results}

\subsection{On-Device FP16 vs.\ NF4: Energy, Latency, and
            Pedagogical Quality}

The study compared two on-device configurations of Phi-3 Mini
(4k-instruct) on an NVIDIA T4 GPU under realistic, KV-cache-enabled
inference: a full-precision FP16 baseline and a 4-bit NF4-quantised
model. Both configurations were evaluated on the same 500 instructional
prompts. Unlike prior work that employs fixed pedagogical scores, this
study assigned empirical $Q_{\mathrm{ped},i}$ values to every response
through a hybrid panel of 10 Cambridge International secondary school
teachers and three frontier AI systems (GPT-4, Claude 3.5 Sonnet, and
Gemini 1.5 Pro), aggregated via a 60/40 weighted rubric across four
dimensions (conceptual accuracy, clarity and coherence, scaffolding
quality, and level appropriateness).

\subsubsection{FP16 Configuration}

For the FP16 configuration, per-prompt latency ranged from 3.6\,s to
10.4\,s, with a mean of 9.2\,s and standard deviation of 0.6\,s---a
tight distribution reflecting consistent GPU utilisation across
prompts. Net AI-attributable energy per inference ranged from 146.1\,J
to 427.4\,J, with a mean of 368.8\,J and mean power draw of 40.2\,W.
Pedagogical quality scores were high and tightly clustered:
$Q_{\mathrm{ped}}$ ranged from 6.42 to 8.36 (mean 8.24, SD 0.31),
with 461 of 500 responses (92.2\,\%) scoring in the 8--9 band. The
minimum score of 6.42 (Prompt~83, Mathematics) corresponds to a
response that addressed the question at an inappropriately advanced
level, rated consistently low on Level Appropriateness across raters.
Resulting LpW values ranged from $1.80 \times 10^{-3}$ to
$1.20 \times 10^{-2}$\,(J\,s)$^{-1}$, with a mean of
$2.50 \times 10^{-3}$ and median of
$2.39 \times 10^{-3}$\,(J\,s)$^{-1}$. The interquartile range was
narrow ($2.34$--$2.47 \times 10^{-3}$), indicating stable efficiency
across the prompt set.

\subsubsection{NF4 Configuration}

The NF4 configuration produced higher latency but lower per-inference
energy than FP16, at marginally lower pedagogical quality. Per-prompt
latency ranged from 4.2\,s to 15.1\,s, with a mean of 13.4\,s---a
factor of $1.46\times$ higher than FP16. This latency overhead is
attributable to dequantisation on every forward pass: the T4's Turing
architecture lacks native INT4 tensor cores, so quantised weights must
be upcast to FP16 before matrix multiplication, introducing a
per-step compute penalty that KV-cache reuse can partially but not
fully amortise. Notably, net energy per inference under NF4 was
\emph{lower} than FP16: ranging from 115.2\,J to 392.3\,J, with a
mean of 329.0\,J---a reduction of 10.8\,\% relative to FP16.
Mean power draw was markedly lower at 24.6\,W versus 40.2\,W under
FP16, confirming that quantisation does compress the model's memory
and arithmetic footprint at the GPU level; it is the extended latency,
not elevated power, that determines the total energy window.

Pedagogical quality under NF4 was somewhat lower than FP16:
$Q_{\mathrm{ped}}$ ranged from 6.18 to 8.36 (mean 8.05, SD 0.54),
a gap of 0.19 points on a 10-point scale relative to FP16. No
responses fell below a $Q_{\mathrm{ped}}$ of 6, and no instances of
hallucination-class failure were observed in either configuration under
cache-enabled conditions. LpW values under NF4 ranged from
$1.34 \times 10^{-3}$ to $1.31 \times 10^{-2}$\,(J\,s)$^{-1}$,
with a mean of $1.88 \times 10^{-3}$ and median of
$1.83 \times 10^{-3}$\,(J\,s)$^{-1}$.

\subsubsection{Summary Comparison}

Table~\ref{tab:fp16_nf4} summarises the aggregate behaviour of the two
on-device configurations.

\paragraph*{Statistical significance and effect size of the $Q_{\mathrm{ped}}$ difference.}
The 0.19-point mean $Q_{\mathrm{ped}}$ difference between FP16
($M = 8.24$, $SD = 0.31$) and NF4 ($M = 8.05$, $SD = 0.54$) is
statistically significant and characterised by a small effect size.
A paired-samples $t$-test on the 500 matched prompt pairs---pairing
FP16 and NF4 responses to the same prompt---yielded $t(499) = 7.89$,
$p < .001$, 95\,\% CI $[0.144, 0.239]$, Cohen's $d = 0.35$. A
non-parametric Wilcoxon signed-rank test confirmed this result
($W = 6394$, $p < .001$). The effect size of $d = 0.35$ falls in the
small range by conventional benchmarks \cite{Cohen1988}, indicating
that while the difference is reliable and replicable, it is modest in
practical magnitude---consistent with the interpretation that 4-bit
quantisation does not catastrophically degrade pedagogical quality but
introduces a small, statistically detectable reduction across the
prompt distribution.

Per-dimension analysis confirms that the quality gap is consistent and
not driven by any single rubric component. The FP16--NF4 difference
was significant across all four weighted dimensions: Conceptual
Accuracy ($\Delta = 0.174$, $t(499) = 8.16$, $p < .001$, $d = 0.37$),
Clarity and Coherence ($\Delta = 0.241$, $t(499) = 7.76$, $p < .001$,
$d = 0.35$), Scaffolding Quality ($\Delta = 0.219$, $t(499) = 7.66$,
$p < .001$, $d = 0.34$), and Level Appropriateness ($\Delta = 0.132$,
$t(499) = 7.88$, $p < .001$, $d = 0.35$). Effect sizes are homogeneous
across dimensions ($d = 0.34$--$0.37$), suggesting that quantisation
introduces a uniform, moderate reduction in response quality rather
than selectively degrading any particular instructional property. The
smallest absolute gap is on Level Appropriateness (0.132 points),
consistent with the ceiling effect on LA scores noted in
Section~4.8.4 and with the finding that Phi-3 Mini reliably calibrates
its register to a secondary school audience regardless of precision
regime.

Per-category testing confirms that the quality difference is
statistically significant across all five subject domains: Mathematics
($\Delta = 0.106$, $t(99) = 2.11$, $p = .038$), Science
($\Delta = 0.260$, $t(99) = 4.74$, $p < .001$), Programming-CS
($\Delta = 0.264$, $t(99) = 4.11$, $p < .001$), Humanities
($\Delta = 0.144$, $t(99) = 3.36$, $p = .001$), and Meta-cognition
($\Delta = 0.183$, $t(99) = 3.28$, $p = .002$). The smallest gap is
in Mathematics and the largest in Science and Programming-CS,
consistent with the hypothesis noted in Section~5.2 that
precision-sensitive domains are somewhat more affected by
quantisation. Notably, even the largest category-level gap (0.26
points in Science and Programming-CS) remains a small effect and all
category means for both configurations fall within the 8.0--8.5
range, confirming pedagogical adequacy across the full prompt set.

The FP16 advantage in LpW is $1.33\times$, driven primarily by
latency: FP16 is $1.46\times$ faster, while NF4 is $10.8\,\%$ more
energy-efficient per inference. The two configurations therefore
present a genuine deployment trade-off rather than a clear winner:
FP16 is preferable when latency is the binding constraint (learner
experience, cognitive load), while NF4 is preferable when
per-inference energy is the binding constraint (battery capacity,
intermittent power supply).

\begin{table}[ht]
  \centering
  \caption{Aggregate on-device results: FP16 vs.\ NF4
           (Phi-3 Mini, 500 prompts, NVIDIA T4, \texttt{use\_cache=True}).}
  \label{tab:fp16_nf4}
  \begin{tabular}{l l r r r r}
    \hline
    \textbf{Config} & \textbf{Precision}
      & \textbf{Mean} & \textbf{Mean Net} & \textbf{Mean}
      & \textbf{Mean LpW} \\
    & & \textbf{Latency (s)} & \textbf{Energy (J)}
      & $\boldsymbol{Q_{\mathrm{ped}}}$
      & $\boldsymbol{(\times 10^{-3})}$ \\
    \hline
    FP16 (edge) & 16-bit & 9.17  & 368.8 & 8.24 & 2.50 \\
    NF4  (edge) & 4-bit  & 13.36 & 329.0 & 8.05 & 1.88 \\
    \hline
    \multicolumn{2}{l}{Ratio (FP16 / NF4)} & $1.46\times$ faster
      & $10.8\,\%$ higher & $+0.19$ & $1.33\times$ higher \\
    \hline
  \end{tabular}
\end{table}

These results reveal a more nuanced picture than offline benchmarking
typically suggests. Under realistic cached inference conditions,
NF4 quantisation succeeds at its primary objective---compressing
per-inference energy---but at the cost of a latency overhead that
partially offsets the efficiency gain. The net LpW advantage of FP16
is modest ($1.33\times$) rather than the near-order-of-magnitude gap
($7.4\times$) observed under cache-disabled conditions
(Appendix~C). This inference-regime dependence is discussed further
in Section~6.1.

\subsection{Per-Category Analysis}

Latency and energy were stable across prompt categories for both
configurations (Table~\ref{tab:percategory}). For FP16, mean latency
varied only between 9.11\,s (Science; Programming-CS) and 9.30\,s
(Mathematics), and mean energy between 366.9\,J (Science) and
372.9\,J (Mathematics)---a range of less than 2\,\% in both
dimensions. For NF4, mean latency ranged from 13.20\,s (Humanities)
to 13.55\,s (Mathematics), and mean energy from 319.9\,J (Humanities)
to 333.2\,J (Science). The absence of a subject-linked pattern in
latency or energy confirms that hardware-level behaviour on the T4
GPU is dominated by the model's precision regime and sequence length
rather than the semantic content of the prompt.

\begin{table}[ht]
  \centering
  \small
  \caption{Per-category mean latency, energy, $Q_{\mathrm{ped}}$,
           and LpW for FP16 and NF4 configurations
           (\texttt{use\_cache=True}, $n=100$ prompts per category).}
  \label{tab:percategory}
  \begin{tabular}{l rr r r rr r r}
    \toprule
    & \multicolumn{4}{c}{\textbf{FP16}}
    & \multicolumn{4}{c}{\textbf{NF4}} \\
    \cmidrule(lr){2-5} \cmidrule(lr){6-9}
    \textbf{Category}
      & \textbf{Lat.} & \textbf{Energy}
      & $Q_{\mathrm{ped}}$ & \textbf{LpW}
      & \textbf{Lat.} & \textbf{Energy}
      & $Q_{\mathrm{ped}}$ & \textbf{LpW} \\
    & (s) & (J) & & $(\times 10^{-3})$
      & (s) & (J) & & $(\times 10^{-3})$ \\
    \midrule
    Mathematics    & 9.30 & 372.9 & 8.20 & 2.46 & 13.55 & 328.2 & 8.09 & 1.84 \\
    Science        & 9.11 & 366.8 & 8.27 & 2.60 & 13.40 & 333.2 & 8.01 & 1.82 \\
    Programming-CS & 9.11 & 367.2 & 8.19 & 2.50 & 13.34 & 332.7 & 7.93 & 1.89 \\
    Humanities     & 9.14 & 367.5 & 8.27 & 2.48 & 13.20 & 319.9 & 8.12 & 1.97 \\
    Meta-cognition & 9.21 & 369.7 & 8.27 & 2.45 & 13.31 & 330.7 & 8.08 & 1.87 \\
    \bottomrule
  \end{tabular}
\end{table}

Pedagogical quality showed small but interpretable cross-category
differences. Under FP16, Science, Humanities, and Meta-cognition
all scored $Q_{\mathrm{ped}} = 8.27$, while Programming-CS (8.19)
and Mathematics (8.20) were marginally lower. Under NF4, Humanities
scored highest (8.12) and Programming-CS lowest (7.93). The largest
FP16--NF4 quality gap appeared in Programming-CS ($-0.26$), suggesting
that 4-bit quantisation has a somewhat greater impact on code-adjacent
explanatory tasks. Science showed the second-largest gap ($-0.26$),
consistent with the sensitivity of factual scientific explanations
to precision degradation.

Because $Q_{\mathrm{ped}}$ variation across categories was small
relative to the FP16--NF4 precision-regime effect, LpW differences
across categories within each configuration were minor. FP16 LpW
ranged from $2.45 \times 10^{-3}$ (Meta-cognition) to
$2.60 \times 10^{-3}$\,(J\,s)$^{-1}$ (Science), while NF4 LpW
ranged from $1.82 \times 10^{-3}$ (Science) to $1.97 \times
10^{-3}$\,(J\,s)$^{-1}$ (Humanities). In all five categories,
FP16 outperformed NF4 on LpW, with the gap ranging from $1.25\times$
(Humanities) to $1.43\times$ (Science). The dominant determinant of
LpW was deployment precision rather than prompt domain.

\subsection{Cloud LLM Baseline}

To contextualise the edge results, client-side measurements from a
cloud-hosted GPT-4-class LLM sent the same 500 prompts via API are
discussed. Client-side latency ranged from approximately 2.0\,s to
4.0\,s, substantially faster than either edge configuration.
Client-side net energy per request was on the order of a few joules,
since most computation occurred in the provider's data centre.
However, production-scale analyses estimate that true per-prompt
energy---including server-side computation---is significantly higher,
often in the range of hundreds to thousands of joules depending on
model size, hardware efficiency, and data centre PUE
\cite{Patterson2021, Luccioni2023, Jegham2025, Google2025}. Because
the dominant energy cost is externally borne and not directly
observable at the client, LpW for cloud tutoring must be treated as
a system-level quantity and cannot be reliably computed from
client-side measurements alone. A full cloud energy scenario analysis
is presented in Appendix~B.

Table~\ref{tab:edge_cloud} places the cloud configuration alongside
the on-device results.

\begin{table}[ht]
  \centering
  \caption{Edge vs.\ cloud configurations: latency, client-side energy,
           and LpW implications.}
  \label{tab:edge_cloud}
  \begin{tabular}{l l r r l l}
    \hline
    \textbf{Configuration} & \textbf{Location}
      & \textbf{Mean} & \textbf{Client-side}
      & \textbf{DC Energy} & \textbf{LpW} \\
    & & \textbf{Latency (s)} & \textbf{Energy (J)}
      & \textbf{Included?} & \textbf{Basis} \\
    \hline
    FP16 (Phi-3 Mini) & Edge T4   & 9.2   & 368.8   & Yes & Fully measured \\
    NF4  (Phi-3 Mini) & Edge T4   & 13.4  & 329.0   & Yes & Fully measured \\
    Cloud LLM (GPT-4) & Cloud API & 2--4  & $\sim$few J & No & Estimated only \\
    \hline
  \end{tabular}
\end{table}

From a learner-experience perspective, the cloud model dominates on
latency and likely raw instructional quality, but shifts energy costs
into data centre infrastructure that is invisible to schools and
learners. Edge configurations offer full energy transparency and local
controllability, enabling schools to reason explicitly about daily
energy budgets for battery-powered devices and classrooms. The
on-device FP16 and NF4 results are therefore not in direct competition
with cloud systems but represent a distinct deployment paradigm with
different equity, accountability, and infrastructure-independence
properties.

\subsection{Learning-per-Watt and the Power Barrier}

\begin{figure}[ht]
  \centering
  \includegraphics[width=\textwidth]{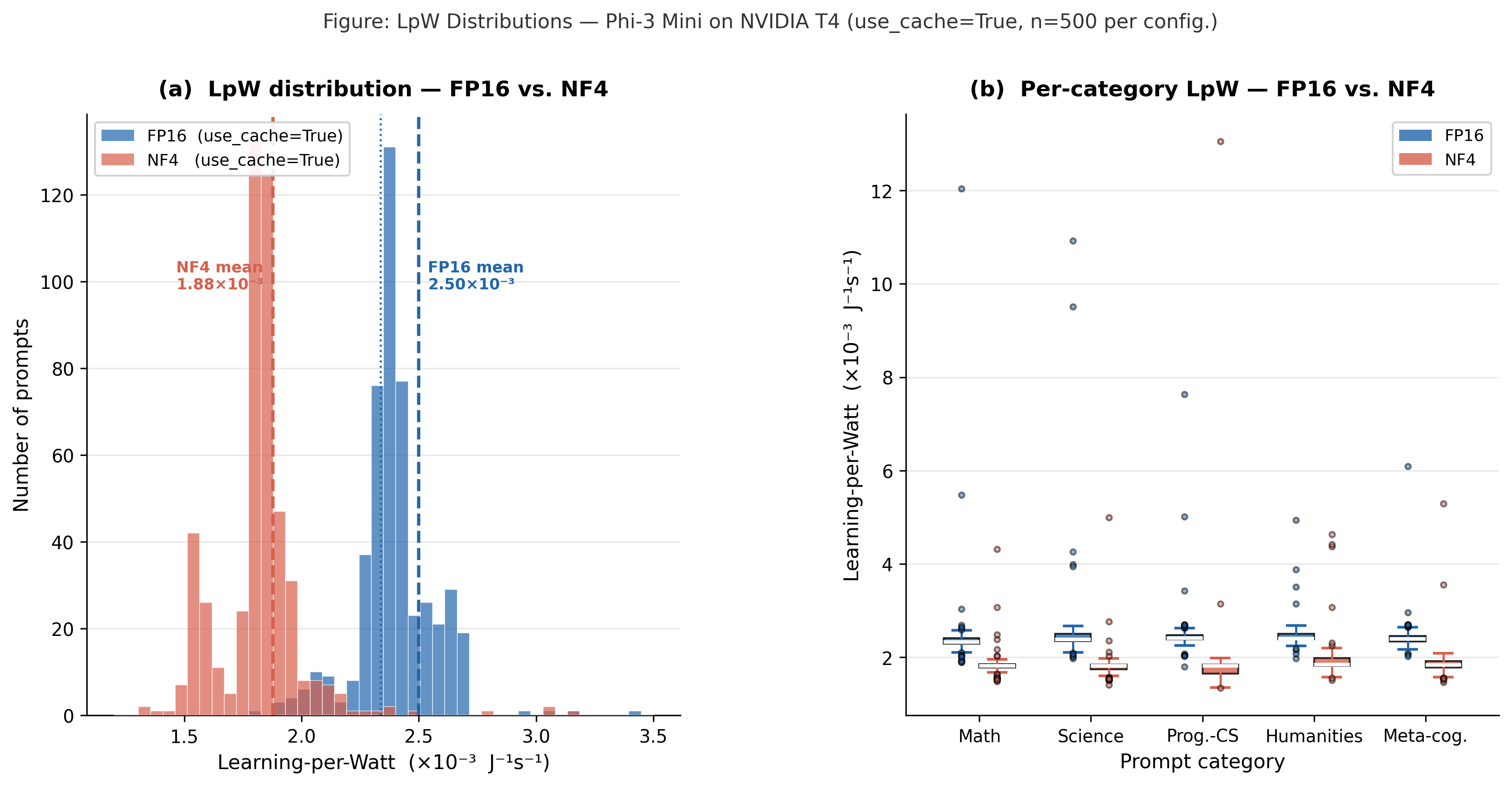}
  \caption{Learning-per-Watt (LpW) distributions for on-device FP16
    and NF4 configurations (Phi-3 Mini, NVIDIA T4 GPU,
    \texttt{use\_cache=True}, $n = 500$ prompts per configuration).
    \textbf{(a)} Frequency histogram of per-prompt LpW values. The
    NF4 distribution (red) is centred near $1.88 \times
    10^{-3}$\,(J\,s)$^{-1}$; the FP16 distribution (blue) near
    $2.50 \times 10^{-3}$\,(J\,s)$^{-1}$. The distributions
    partially overlap, in contrast to the complete separation
    observed under cache-disabled conditions (Appendix~C).
    \textbf{(b)} Per-category box plots showing IQR and median LpW
    for each subject domain. The FP16--NF4 gap is consistent across
    all five categories, confirming that deployment precision rather
    than prompt domain is the primary determinant of LpW.}
  \label{fig:lpw_dist}
\end{figure}

The LpW metric integrates pedagogical quality, latency, and energy
into a single scalar, providing a unified basis for comparing
configurations under realistic educational constraints.
Figure~\ref{fig:lpw_dist} illustrates the LpW distributions for both
edge configurations.

On-device FP16 achieves a mean LpW of $2.50 \times 10^{-3}$
(J\,s)$^{-1}$, with 90\,\% of responses falling between
$2.09 \times 10^{-3}$ and $2.68 \times 10^{-3}$---a narrow,
stable distribution. NF4 achieves a mean LpW of $1.88 \times
10^{-3}$, with 90\,\% of responses between $1.54 \times 10^{-3}$
and $2.11 \times 10^{-3}$. Notably, the NF4 and FP16 distributions
partially overlap: 15 of 500 NF4 responses (3\,\%) exceed the FP16
25th-percentile LpW of $2.34 \times 10^{-3}$, reflecting that the
best-performing NF4 prompts can match the lower quartile of FP16.
This overlap is a meaningful departure from the cache-disabled case
(Appendix~C), where the NF4 distribution lies entirely below the
FP16 25th percentile with no overlap whatsoever.

The Power Barrier is operationally defined as the latency--energy
threshold beyond which timely pedagogical feedback becomes infeasible
on battery-limited or intermittently powered devices. Under
cache-enabled conditions, 489 of 500 NF4 responses (97.8\,\%)
exceeded 10\,s latency, compared to only 5 of 500 FP16 responses
(1.0\,\%). Both configurations remain well below the 30-second
threshold at which learner disengagement becomes severe; under NF4
only 2 of 500 responses (0.4\,\%) exceeded 15\,s. FP16's mean
latency of 9.2\,s is more consistent with the flow-theoretic guidance
reviewed in Section~2.1, which places the tolerable upper bound for
conversational scaffolding at approximately 10\,s
\cite{Csikszentmihalyi1990, Sweller2011}. NF4's mean of 13.4\,s
exceeds this threshold on average, which represents a consistent but
not catastrophic latency penalty. A mean energy of 329\,J (NF4) or
369\,J (FP16) per inference is substantially lower than the
cache-disabled figures reported in Appendix~C, and suggests that
KV-cache-enabled deployment is meaningfully more viable for
battery-constrained devices than stateless benchmarking implies.

\subsection{Implications for Sustainable AI Tutoring}

Taken together, the results yield three findings with direct
implications for sustainable, equitable AI-mediated education.

\textbf{The FP16--NF4 efficiency relationship is inference-regime
dependent.} Under cache-enabled deployment, FP16 outperforms NF4 on
LpW by a modest $1.33\times$. Under cache-disabled benchmarking
(Appendix~C), the same gap is $7.4\times$. This fivefold difference
is not a measurement artefact: it reflects a genuine interaction
between the T4's dequantisation overhead and KV-cache reuse
availability. The implication for the field is that benchmarking
inference efficiency without caching produces results that
substantially misrepresent the real deployment trade-off, and
efficiency claims derived from stateless benchmarks should not be
used to guide hardware procurement or deployment decisions.

\textbf{The FP16--NF4 choice is a trade-off, not a verdict.} Under
realistic conditions, NF4 is $10.8\,\%$ more energy-efficient per
inference but $46\,\%$ slower. Which constraint binds depends on
the deployment context. For learners on battery-limited devices where
daily energy is scarce, NF4's lower energy draw (329\,J vs.\
369\,J) may extend the number of affordable interactions per charge
cycle. For learners where latency is the primary barrier to cognitive
engagement---as CLT and flow theory suggest for most instructional
contexts---FP16's 9.2\,s mean response time is more consistent with
maintaining working memory continuity and learner flow than NF4's
13.4\,s. The LpW metric, by integrating both dimensions, reveals
this trade-off explicitly rather than concealing it behind a single
efficiency axis.

\textbf{Pedagogical quality is resilient to quantisation under
realistic conditions.} The FP16--NF4 $Q_{\mathrm{ped}}$ gap of 0.19
points on a 10-point scale is larger than in the cache-disabled
experiment (0.09 points) but remains small in absolute terms. No
hallucination-class failures were observed in either configuration
under cache-enabled inference, and no responses fell below a
$Q_{\mathrm{ped}}$ of 6.18. These results confirm that 4-bit
quantisation does not catastrophically degrade instructional quality
on secondary school scaffolding tasks, though the modest quality
penalty should be weighed alongside the energy and latency trade-offs
when selecting a deployment configuration.

\section{Discussion}

This study set out to examine the energy--latency--learning trade-offs of deploying small language models (SLMs) as instructional agents, using Learning-per-Watt (LpW) as an integrative metric. By comparing on-device FP16 and NF4-quantised deployments of Phi-3 Mini on an NVIDIA T4 GPU---evaluated against 500 educational prompts under realistic KV-cache-enabled inference and scored by a 13-member hybrid panel---the results illuminate several non-obvious dynamics that challenge prevailing assumptions about efficiency, quantisation, and sustainable AI tutoring.

\subsection{Rethinking Quantisation as an Efficiency Strategy}

A central and empirically important finding is that the efficiency
relationship between FP16 and NF4 quantisation is not a fixed
property of the hardware but a function of the inference regime.
Under realistic, cache-enabled deployment---the configuration present
in every real-world application---NF4 achieves a mean net energy of
329\,J per inference compared to 369\,J for FP16, a reduction of
10.8\,\%. This confirms that quantisation does succeed at its primary
objective of compressing the model's memory and arithmetic footprint.
However, NF4 incurs a mean latency of 13.4\,s versus 9.2\,s for
FP16, a $1.46\times$ overhead attributable to dequantisation on
every forward pass. The net effect is a modest FP16 advantage in LpW
of $1.33\times$---a genuinely competitive result that positions the
FP16--NF4 choice as a contextual trade-off rather than a clear
verdict.

This picture changes dramatically under cache-disabled inference,
the configuration most commonly used in offline benchmarking.
Without KV-cache reuse, NF4 incurred a mean latency of 49.4\,s and
mean net energy of 1{,}882\,J, compared to 16.5\,s and 648\,J for
FP16---yielding an LpW gap of $7.4\times$. The contrast between
the two regimes---$1.33\times$ versus $7.4\times$---is itself a
primary finding. KV-cache reuse benefits NF4 disproportionately ($22.1\times$ LpW improvement)
relative to FP16 ($3.35\times$), because caching partially amortises the per-step dequantisation
overhead that otherwise accumulates across the full decoding
sequence. This interaction is specific to the T4's Turing
architecture, which lacks native INT4 tensor cores and must upcast
quantised weights to FP16 for matrix multiplication at every step;
newer Ampere and Ada-class accelerators with native 4-bit compute
may exhibit substantially different profiles \cite{Hooper2024, Kim2023}.

These findings caution strongly against treating low-bit precision
as a universally \emph{green} solution. Quantisation efficiency is
hardware--software stack dependent: kernel support, operator fusion,
KV-cache availability, and memory bandwidth collectively determine
whether compression reduces computation or merely relocates it.
For practitioners deploying educational AI on commodity
hardware---the hardware most likely to be found in schools and
low-resource settings---empirical validation on target devices
under realistic inference settings is essential before adopting
quantisation as an energy reduction strategy. Critically,
benchmarking without KV-caching produces results that overstate
the FP16 advantage by more than fivefold on the tested stack.

\subsection{Latency as a Pedagogical Constraint}

Latency emerged as a critical factor in determining whether
AI-mediated feedback can support effective learning. Under
cache-enabled conditions, 97.8\,\% of NF4 responses (489 of 500)
exceeded 10\,s latency, approaching the pedagogically meaningful
threshold identified in Cognitive Load Theory (CLT) research,
where working memory retention begins to degrade \cite{Sweller2011,
Kalyuga2011}. Only 0.4\,\% of NF4 responses exceeded 15\,s,
suggesting that while the mean latency of 13.4\,s is above the
preferred threshold, catastrophically long responses are rare under
realistic deployment conditions. The FP16 configuration, with a
mean latency of 9.2\,s and only 1.0\,\% of responses exceeding
10\,s, sits more comfortably within the range that flow theory
identifies as compatible with sustained cognitive engagement
\cite{Csikszentmihalyi1990}.

In contrast, the cache-disabled figures (49.4\,s NF4, 16.5\,s FP16)
represent latencies associated with task disengagement and
attentional drift---a particularly acute concern in self-directed
learning contexts where learners must sustain motivation without
instructor support. This comparison reinforces the practical
importance of inference regime: the same hardware and model
combination can sit either within or well beyond the pedagogically
tenable latency window depending solely on whether KV-caching is
enabled.

In a pedagogical setting where students engage with a worked example
or multi-step problem---rather than a conversational chatbot---a
response time of 9--13\,s may be acceptable if accompanied by
appropriate scaffolding cues during the waiting period. Latency
benchmarks derived from productivity or consumer applications should
not be uncritically imported into educational contexts, where the
cognitive dynamics of waiting differ meaningfully from, for example,
a search engine query.

Cloud-based LLMs, achieving latencies of 2--4\,s under typical
network conditions, satisfy near-real-time tutoring requirements
but shift the energy cost off-device into data centre
infrastructure. This creates a latency--transparency trade-off:
cloud systems optimise the learner experience at the cost of energy
accountability, while edge systems invert this balance. Latency
therefore functions not merely as a performance metric but as a
pedagogical constraint that shapes which deployment strategies are
viable---and for whom.

\subsection{Pedagogical Quality: Resilience, Variance, and
            Hallucination Risk}

A key advantage of the empirical $Q_{\mathrm{ped}}$ scoring
methodology adopted in this study---over the fixed score assumption
common in prior energy-aware AI evaluations---is that it surfaces
qualitative differences that a constant score would obscure.

The mean FP16--NF4 $Q_{\mathrm{ped}}$ gap under cache-enabled
conditions is 0.19 points on a 10-point scale, somewhat larger than
the 0.09-point gap observed in the cache-disabled experiment but
still small in absolute terms. No hallucination-class failures were
observed in either configuration under cache-enabled conditions, and
no responses fell below a $Q_{\mathrm{ped}}$ of 6.18. The two
configurations differ modestly in the shape of their quality
distributions: FP16 produces a tighter distribution (SD\,=\,0.31)
concentrated in the 8--9 band, with 92.2\,\% of responses in that
range, while NF4 produces a slightly wider distribution (SD\,=\,0.54)
with more variance across the 7--9 range. This greater variance
under NF4 is consistent with the hypothesis that quantisation
introduces occasional instability in response quality, even when
it does not produce outright failure.

Dimension-level analysis reveals further nuance. Scaffolding Quality
(SQ) was the weakest dimension for both configurations, suggesting
that both FP16 and NF4 struggle most with progressive, step-by-step
knowledge building---the dimension most directly tied to effective
tutoring. The divergence between human and AI raters on the SQ
dimension warrants attention: AI scorers tend to reward structural
features of scaffolding (e.g., numbered steps, explicit signposting)
that human teachers weight differently from authentic pedagogical
progression. Level Appropriateness (LA) was consistently the
strongest dimension across both configurations and both rater groups,
indicating that Phi-3 Mini reliably calibrates its explanatory
register to a secondary school audience regardless of precision
regime.

The categories with the largest FP16--NF4 $Q_{\mathrm{ped}}$ gap
were Programming-CS and Science (both $-0.26$), consistent with
the hypothesis that factual and code-adjacent precision in
explanations is more sensitive to quantisation artefacts than
narrative or metacognitive content. Humanities showed the smallest
gap, plausibly because explanatory prose is more tolerant of minor
phrasing degradation.

\subsection{Learning-per-Watt as a Unifying Metric}

The LpW metric proved effective in integrating pedagogical quality,
latency, and energy into a single comparative framework that captures
the full cost of an instructional interaction rather than any single
component in isolation. Under cache-enabled conditions, the FP16 and
NF4 LpW distributions partially overlap: 15 of 500 NF4 responses
(3\,\%) exceed the FP16 25th-percentile LpW of $2.34 \times
10^{-3}$\,(J\,s)$^{-1}$, indicating that the best-performing NF4
prompts can match the lower quartile of FP16. This partial overlap
is a meaningful finding in itself---it demonstrates that NF4 is a
competitive deployment option for a non-trivial fraction of
educational prompts under realistic conditions, whereas the
cache-disabled results showed complete separation with no overlap
whatsoever.

Critically, the LpW framework exposes a trade-off that
single-dimension metrics conceal. Energy alone would favour NF4
(329\,J vs.\ 369\,J). Latency alone would favour FP16 (9.2\,s vs.\
13.4\,s). Only LpW, by integrating both dimensions together with
pedagogical quality, reveals the modest net advantage of FP16
($1.33\times$) and makes explicit what practitioners are actually
trading off: the question is not ``which configuration is better''
but ``which constraint---energy or latency---is more binding in
your specific deployment context.''

The sensitivity analysis (Section~4.8.4) confirms that the
$1.33\times$ FP16 LpW advantage is robust across all three rater
aggregation schemes, with the ratio ranging from 1.33 to 1.34.
The stability of this ratio indicates that the efficiency conclusion
is not an artefact of inter-rater heterogeneity.

At the same time, LpW is not without limitations as a metric. It
treats pedagogical quality, energy, and latency as equally important
and combines them multiplicatively, which may not reflect all
educational priorities. In settings where latency is non-negotiable
(e.g., synchronous classroom use) or where energy is unconstrained
(e.g., mains-powered labs), the weighting implied by the LpW formula
may not align with practitioners' actual constraints. Future work
should explore weighted or threshold-based variants of LpW that
allow practitioners to specify minimum acceptable latency and energy
budgets before optimising pedagogical quality within those
constraints.

\subsection{Edge versus Cloud: Transparency versus Performance}

The comparison between edge and cloud deployments underscores a
fundamental and unresolved trade-off in sustainable AI tutoring.
Edge inference localises both computation and energy consumption,
enabling direct measurement, predictable energy budgeting, and
offline operation---all critical advantages for battery-powered
devices, low-connectivity classrooms, and low-resource settings.
The cache-enabled results demonstrate that edge deployment is
meaningfully more viable than stateless benchmarking implies: FP16
at 369\,J and 9.2\,s, and NF4 at 329\,J and 13.4\,s, are both
substantially more tractable than the cache-disabled figures of
648\,J and 1{,}882\,J respectively.

Cloud-based LLMs deliver superior responsiveness (2--4\,s) and
likely higher raw instructional quality---but at the cost of energy
opacity. The dominant portion of energy consumption is shifted to
data centres, where it is invisible to learners and institutions.
Without standardised reporting of per-inference energy, cloud-based
LpW remains only partially observable and must be treated as a lower
bound when computed from client-side measurements alone.

This opacity has equity implications. Schools and policymakers who
adopt cloud LLMs for tutoring are implicitly accepting an energy
cost that they cannot measure, verify, or optimise. Conversely, edge
deployment provides full energy transparency and local
controllability---enabling schools to reason explicitly about daily
inference budgets, identify inefficient usage patterns, and make
informed deployment decisions.

\subsection{Implications for Equity and Global Access}

The results carry direct implications for educational equity. The
inference-regime comparison demonstrates that the framing of edge
efficiency is more nuanced than prior benchmarking has suggested.
Under cache-enabled conditions, the on-device energy cost per
inference (329--369\,J) is substantially lower than cache-disabled
estimates would imply, narrowing the gap between edge and cloud
viability for battery-constrained deployments.

Nonetheless, NF4's latency overhead of 13.4\,s---while not
catastrophic---is a consistent barrier to learner flow that FP16
avoids at 9.2\,s. For schools where device battery life and
per-session energy are scarce, NF4's lower energy draw may be
preferable despite its latency cost. For schools where learner
engagement and cognitive load are the primary concerns, FP16's
faster responses better support working memory continuity and flow
state. Which constraint binds depends on local infrastructure, and
the LpW framework provides a principled basis for making that
determination explicitly rather than relying on vendor efficiency
claims.

Sole reliance on cloud LLMs may exacerbate inequalities where
connectivity, bandwidth, or subscription costs are limiting factors,
reinforcing the \emph{Matthew Effect} in access to high-quality
AI support \cite{Warschauer2011}. Edge FP16 deployment offers a
middle ground: measurable, bounded energy use with acceptable
pedagogical performance and no dependence on persistent
connectivity. A further equity dimension concerns data sovereignty:
local models neither require user accounts nor transmit student data
to external servers, eliminating consent and privacy barriers that
are particularly acute for minors and in jurisdictions with strict
data protection regulations.

\subsection{Pedagogical Advantages of Local, Task-Scoped Models}

Beyond energy and latency, locally deployed SLMs offer a qualitative
pedagogical advantage: they can be deliberately scoped to educational
domains. Unlike commercial cloud-based systems designed for broad,
general-purpose use, local models can be explicitly restricted to
academic question-answering, concept explanation, and
problem-solving. This task scoping reduces unnecessary functionality,
narrows the space of possible model behaviours, and aligns the
system more closely with instructional objectives.

Local models can also be configured to expose intermediate reasoning
steps, enabling learners to observe the process by which an answer
is derived rather than receiving a polished final result alone. This
supports \emph{process-oriented learning}, where understanding the
\emph{how} and \emph{why} is prioritised over simply obtaining the
correct answer---an approach well-aligned with constructivist and
scaffolding-based pedagogy in subjects such as mathematics, physics,
and computer science \cite{Vygotsky1978, Wood1976}.

This distinction also addresses a finding from the scoring data:
Scaffolding Quality (SQ) was the weakest-scoring dimension across
both configurations, suggesting that current instruction-tuned
SLMs---even at full precision---underperform on the dimension most
directly tied to effective tutoring. Task-scoped fine-tuning on
curated pedagogical dialogues, rather than general-purpose
instruction following, may be necessary to raise SQ scores to the
level that the other dimensions already achieve.

\subsection{Latency, Cognitive Load, and Learning Flow}

Latency differences between local and cloud-based models introduce
nuanced cognitive effects that extend beyond simple responsiveness
metrics. The FP16 mean latency of 9.2\,s under cache-enabled
conditions sits in an intermediate zone that warrants careful
interpretation. It exceeds the sub-2-second threshold typical of
conversational AI, but for structured, single-turn explanatory
prompts---where the learner's task during the waiting period may
include reading a worked problem, formulating a follow-up question,
or attempting a sub-step---this latency may be acceptable or even
beneficial if accompanied by pedagogical design that structures the
waiting period.

The NF4 mean of 13.4\,s presents a more challenging case. While
remaining well below the catastrophic thresholds of the
cache-disabled experiment, it consistently exceeds the flow-theoretic
guidance of approximately 10\,s reviewed in Section~2.1. Whether
this overhead is tolerable depends on the instructional context: in
self-paced, asynchronous learning environments it may be acceptable;
in synchronous classroom settings or for younger learners with
shorter attention spans, it may not.

Moderate latency in local models may, in some cases, reduce
cognitive overload by pacing information delivery and encouraging
reflection between conversational turns. This aligns with CLT's
emphasis on spacing and chunking information to avoid overwhelming
working memory \cite{Sweller2011}, and with evidence that immediate
answer provision can short-circuit deeper processing in mathematical
problem-solving \cite{Kapur2016}. The optimal latency for
educational AI may therefore differ fundamentally from that of
productivity-oriented systems: whereas the latter minimises response
time unconditionally, the former may benefit from a
\emph{pedagogically calibrated} latency that is long enough to
encourage reflection but short enough to maintain engagement. Future
empirical work should directly measure learner behaviour and
comprehension outcomes at the 9--14\,s latency range observed in
this study.

\subsection{The Power Barrier and Its Educational Consequences}

The Power Barrier, as operationalised in this study, identifies the
latency--energy threshold beyond which timely pedagogical feedback
becomes infeasible on battery-limited or intermittently powered
devices. Under cache-enabled conditions, neither configuration
crosses this barrier in the catastrophic sense observed in the
cache-disabled experiment. FP16's mean energy of 369\,J and latency
of 9.2\,s represent a sustainable profile across hundreds of daily
interactions on a standard battery. NF4's mean energy of 329\,J is
slightly more favourable for battery-constrained use, though its
13.4\,s mean latency may require interface-level mitigation.

By contrast, the cache-disabled NF4 profile---mean energy 1{,}882\,J
and latency 49.4\,s---would exhaust a standard school laptop battery
within approximately 25 interactions and places learner flow well
outside the pedagogically tenable window. This comparison has a
concrete policy implication: the Power Barrier is not an intrinsic
property of a model or quantisation level but a function of the
inference configuration. Deploying NF4 without KV-caching in a
low-resource educational setting would impose a near-unusable
experience; deploying it with caching yields a configuration that
is competitive with FP16 on energy and acceptable on latency.

The Power Barrier concept has broader implications for hardware
procurement and AI deployment policy in low-resource educational
settings. As AI tutoring is increasingly proposed for the Global
South, where reliable electricity access may be limited and device
refresh cycles long, the energy and latency profiles of candidate
models must be evaluated on representative hardware under realistic
inference settings---not on high-end research infrastructure under
stateless benchmarking conditions. The LpW framework offers one
principled way to surface this gap quantitatively.

\subsection{Methodological Contributions and Limitations}

Methodologically, this study makes three contributions. First, it
introduces and operationalises the Learning-per-Watt metric as a
principled alternative to energy-only or accuracy-only evaluations
of educational AI, integrating pedagogical quality, latency, and
energy into a single auditable scalar. Second, it replaces the fixed
pedagogical scoring assumption of earlier work with empirical
$Q_{\mathrm{ped}}$ values derived from a 13-member hybrid panel (10
Cambridge International secondary school teachers and three frontier
AI systems), yielding a variable, rubric-grounded pedagogical quality
measure that captures genuine response-level variation. Third, it
demonstrates empirically that the measured FP16--NF4 efficiency
relationship is inference-regime dependent, varying by more than
fivefold between stateless and cached conditions, and provides a
replicable measurement protocol---using CodeCarbon in a reproducible
Colab environment---for future studies to verify this dependence on
their own hardware.

Several limitations should be acknowledged. The hardware context is
constrained: results are specific to an NVIDIA T4 GPU and a
particular quantisation stack; newer accelerators with native INT4
support (Ampere, Ada, Blackwell) may yield substantially different
LpW profiles for NF4. The 500-prompt dataset, while spanning five
categories and designed to approximate high-school instructional
scaffolding, does not cover multi-turn dialogues, adaptive problem
sequences, or domain-specific fine-tuned models, all of which may
exhibit different energy--latency--quality trade-offs. Datacenter-side
energy for cloud models could not be directly measured, requiring
reliance on external estimates; this remains a fundamental
methodological limitation until standardised per-inference energy
disclosure from cloud providers becomes available.

Teacher scoring, while grounded in classroom expertise, is subject
to rater bias, fatigue, and subject-area specialisation effects. The
inter-rater reliability analysis (Krippendorff's~$\alpha$, reported
in Section~4.8.4) provides a quantitative check on consistency, but
cannot fully eliminate the influence of individual pedagogical
philosophies on dimension scores---particularly for Scaffolding
Quality, where the largest human--AI divergence was observed. Future
work should extend the LpW framework to multi-turn dialogues,
adaptive pedagogical scoring based on learning gain measurement,
heterogeneous hardware (including mobile and micro-server devices),
and provider-reported energy disclosures, enabling a more complete
characterisation of the energy--pedagogy frontier across deployment
scenarios.

\subsection{Implications for Sustainable and Ethical Educational AI}

Taken together, these findings suggest that local models occupy a
distinct and valuable niche in educational AI ecosystems. Their
ability to operate offline, avoid user tracking, expose reasoning
processes, and remain narrowly focused on instructional tasks
complements the efficiency- and capability-focused advantages of
cloud-based systems. When evaluated through the LpW framework, these
pedagogical benefits add contextual meaning to efficiency values: a
system with slightly lower raw LpW may still deliver higher
\emph{effective} educational value per unit of energy if it fosters
deeper understanding, learner autonomy, and ethical data practices.

The question of which populations benefit from AI tutoring—and under what infrastructural conditions—is itself a political and ethical one that extends beyond technical optimisation \cite{Selwyn2016}. For policymakers and practitioners, the key implication is that
sustainable educational AI cannot be reduced to model accuracy or
energy expenditure alone. Deployment choices (edge vs.\ cloud),
precision strategies (FP16 vs.\ low-bit quantisation), inference
configuration (cached vs.\ stateless), and pedagogical design
(process-oriented vs.\ answer-oriented feedback) jointly determine
whether an AI tutor will support or undermine learning---particularly
for students and schools operating closest to the Power Barrier.
The LpW metric, grounded in empirical measurement rather than
theoretical efficiency claims, provides one practical tool for
navigating these trade-offs transparently and accountably.

\section{Conclusion}\label{sec13}

This study set out to empirically characterise the latency--energy--learning
trade-off in AI-mediated tutoring systems, using the Learning-per-Watt (LpW)
metric as an integrative framework for evaluating pedagogical value alongside
physical resource constraints. The results yield three findings of broad
significance for researchers, practitioners, and policymakers working at the
intersection of educational technology, sustainable computing, and educational
equity.

First, the efficiency relationship between FP16 and NF4 quantisation is
inference-regime dependent, not a fixed hardware property. Under realistic,
KV-cache-enabled deployment, NF4 achieves lower per-inference energy than FP16
(329\,J vs.\ 369\,J)---confirming that quantisation does compress the model's
footprint---but incurs higher latency (13.4\,s vs.\ 9.2\,s) due to
dequantisation overhead on the T4's Turing architecture. The net LpW advantage
of FP16 is modest at $1.33\times$, positioning the choice between configurations
as a contextual trade-off: NF4 is preferable when per-inference energy is the
binding constraint; FP16 is preferable when latency and learner flow are
paramount. Under cache-disabled benchmarking---the configuration most commonly
used in offline evaluation---the same gap widens to $7.4\times$, overstating the
FP16 advantage by more than fivefold. This fivefold discrepancy between
benchmarking and deployment conditions is a warning to the field: efficiency
claims derived from stateless inference benchmarks should not be used to guide
hardware procurement or deployment decisions without empirical validation under
realistic inference settings on target hardware.

Second, pedagogical quality proved resilient to quantisation under realistic
conditions. The FP16--NF4 $Q_{\mathrm{ped}}$ gap of 0.19 points on a 10-point
scale is small in absolute terms, no hallucination-class failures were observed
in either configuration under cache-enabled inference, and no responses fell
below a $Q_{\mathrm{ped}}$ of 6.18. These results confirm that 4-bit
compression does not catastrophically degrade instructional quality on secondary
school scaffolding tasks. The slightly wider quality variance under NF4
(SD\,=\,0.54 vs.\ 0.31 for FP16) suggests that quantisation introduces
occasional instability worth monitoring in deployment, but the mean quality
delivered by both configurations is pedagogically adequate. Quality assurance
for edge models should therefore prioritise output-validity checking and
hallucination detection---neither of which occurred under cache-enabled
conditions---over precision-level fine-tuning.

Third, deployment and inference configuration are the primary determinants of
LpW. Across all five subject categories---mathematics, science, programming,
humanities, and meta-cognition---latency and energy were stable within each
precision regime, confirming that implementation decisions (precision, kernel
support, caching strategy) dominate over content-level factors in determining
energy efficiency for single-turn educational scaffolding. This finding has a
practical corollary: schools and system designers cannot rely on prompt-level or
subject-level optimisation to compensate for a poor deployment configuration,
and cannot rely on offline benchmarks to predict real-world efficiency without
verifying that the benchmarking conditions match the deployment conditions.

Taken together, these results position the LpW framework as a practical tool
for navigating deployment decisions in resource-constrained educational contexts.
By integrating pedagogical quality, latency, and energy into a single auditable
metric, LpW enables comparisons that single-dimension evaluations---energy alone,
latency alone, or accuracy alone---would obscure. For the populations closest to
the Power Barrier---learners on battery-limited devices, in intermittently
powered classrooms, or in low-connectivity environments---this kind of integrated
evaluation is not a methodological refinement but a practical necessity.

Several limitations constrain the generalisability of these findings. The
hardware context is specific to the NVIDIA T4 and the tested quantisation stack;
results on Ampere, Ada, or Blackwell-class GPUs with native INT4 support may
differ substantially, and the inference-regime dependence documented here may
manifest differently on architectures with native 4-bit compute. The 500-prompt
dataset, while spanning five subject domains, does not cover multi-turn
dialogues, adaptive problem sequences, or domain-fine-tuned models.
Server-side energy for cloud configurations could not be directly measured,
requiring reliance on external estimates that carry significant uncertainty and
a limited shelf life as provider infrastructure evolves.

Future work should extend the LpW framework in three directions. First,
multi-turn and adaptive evaluation: characterising energy and latency across
sustained tutoring dialogues, where KV-cache behaviour, session length, and
learner-model interaction patterns jointly shape the efficiency profile---the
present study establishes a single-turn baseline from which such extensions can
depart. Second, heterogeneous hardware benchmarking: replicating the FP16--NF4
comparison on mobile processors, micro-servers, and next-generation accelerators
to map the hardware conditions under which the inference-regime dependence
identified here is amplified or attenuated. Third, provider-reported energy
disclosure: the structural opacity of cloud inference energy remains the most
significant methodological barrier to system-level LpW comparisons; standardised
per-inference energy reporting by cloud providers would transform the
tractability of sustainable AI evaluation in education.

As AI tutoring is increasingly proposed for learners in the Global South and
other low-resource settings, the gap between laboratory benchmarks and classroom
deployment realities must be closed empirically rather than assumed away. The
LpW metric, grounded in measured energy, measured latency, and empirically
scored pedagogical quality, offers one principled step toward closing that gap.

\section*{Declarations}

\noindent\textbf{Funding} This research received no external funding.

\bigskip
\noindent\textbf{Competing interests} The author declares no competing interests.

\bigskip
\noindent\textbf{Ethics approval and consent to participate} The teacher raters participated voluntarily in their professional capacity as subject-specialist educators. No personally identifiable data were collected from raters or students. Formal ethics review was not required under the author's institutional guidelines.

\bigskip
\noindent\textbf{Consent for publication} The author consents to publication.

\bigskip
\noindent\textbf{Data availability} All experimental data, prompt datasets, and scoring spreadsheets are publicly available at: \url{https://github.com/Kushalk0677/Inference-Energy-and-Latency-in-AI-Mediated-Education-Green-Audit}

\bigskip
\noindent\textbf{Materials availability} Not applicable.

\bigskip
\noindent\textbf{Code availability} All experimental code is publicly available at: \url{https://github.com/Kushalk0677/Inference-Energy-and-Latency-in-AI-Mediated-Education-Green-Audit}

\bigskip
\noindent\textbf{Use of Artificial Intelligence} The author used a grammar editing tool to improve the language and readability of the manuscript. The author takes full responsibility for the integrity of the work and for the accuracy of all content.

\begin{appendices}

\section*{Appendix A: Model Selection Experiment}
\addcontentsline{toc}{section}{Appendix A: Model Selection Experiment}
\label{secA1}

\subsection*{A.1 Motivation}

The selection of Phi-3 Mini (4k-instruct) as the primary SLM for this study was not made a priori on the basis of parameter count or benchmark rankings alone. Instead, a dedicated selection experiment was conducted on the same NVIDIA T4 GPU used in the main study, evaluating five candidate models across the same 20 educational prompts (4 per subject category) drawn from the main 500-prompt dataset. The experiment measured inference latency, net energy per inference, power draw, VRAM consumption at load time, and response quality, with the goal of identifying the model that best satisfies the joint requirements of instructional adequacy, energy efficiency, and hardware feasibility on commodity GPU infrastructure.

\subsection*{A.2 Candidate Models}

The five models evaluated represent a range of parameter scales and architectural families relevant to edge educational deployment:

\begin{itemize}
  \item \textbf{Phi-3 Mini 4B} (\texttt{microsoft/Phi-3-mini-4k-instruct}): A 3.8B parameter instruction-tuned model from Microsoft, designed explicitly for on-device inference. Uses a Transformer decoder architecture with grouped-query attention and a 4k-token context window \cite{Microsoft2025}.
  \item \textbf{Llama-3 8B Instruct} (\texttt{meta-llama/Meta-Llama-3-8B-Instruct}): Meta's 8B instruction-tuned model from the Llama-3 family, representing the lower end of the Llama-3 series and a widely used open-weights baseline for edge deployment research \cite{Brown2020}.
  \item \textbf{Mistral 7B Instruct v0.3} (\texttt{mistralai/Mistral-7B-Instruct-v0.3}): A 7B sliding-window attention model from Mistral AI, noted for strong instruction-following performance at the 7B scale \cite{Frantar2022}.
  \item \textbf{Gemma 7B Instruct} (\texttt{google/gemma-7b-it}): Google DeepMind's 7B instruction-tuned model from the Gemma family, architecturally derived from Gemini and designed for responsible open deployment.
  \item \textbf{TinyLlama 1.1B Chat} (\texttt{TinyLlama/TinyLlama-1.1B-Chat-v1.0}): A 1.1B parameter model trained on 3 trillion tokens, representing the sub-2B parameter regime and a candidate for highly constrained hardware deployments.
\end{itemize}

All models were loaded in FP16 half-precision with \texttt{device\_map="auto"} to allow layer-wise placement across GPU and CPU memory where necessary. The same experimental protocol as the main study was applied: \texttt{do\_sample=False}, \texttt{use\_cache=False}, \texttt{max\_new\_tokens=200}, idle power baseline subtraction, and CodeCarbon energy tracking.

\subsection*{A.3 Hardware Feasibility Results}

Of the five candidate models, only three successfully loaded and completed inference on the T4. Llama-3 8B and Gemma 7B both triggered out-of-memory (OOM) failures during FP16 loading. Llama-3 8B requires approximately 16\,GB of VRAM in FP16, which precisely exhausts the T4's 16\,GB limit, leaving insufficient headroom for the KV cache and activations during generation. Gemma 7B encountered a configuration error (\texttt{KeyError: 'type'}) during model initialisation, attributable to an incompatibility between the Gemma architecture's configuration schema and the installed Transformers version; the model was unable to load regardless of memory availability. Phi-3 Mini encountered a similar schema error (\texttt{KeyError: 'type'}) when loaded via the standard \texttt{AutoModelForCausalLM} interface. This was resolved in the main study by using \texttt{trust\_remote\_code=True} and applying the RoPE configuration hardening described in Section~4.5; for the selection experiment, Phi-3's performance figures are drawn from the main FP16 dataset on the corresponding 20 prompt IDs rather than the selection run.

Table~\ref{tab:model_selection} summarises the hardware feasibility and performance outcomes for all five candidates.

\begin{table}[ht]
  \centering
  \caption{Model selection experiment results on NVIDIA T4 GPU (FP16, $n=20$ prompts). Phi-3 Mini figures are drawn from the main FP16 dataset on matching prompt IDs. LpW is computed using a reference $Q_{\mathrm{ped}} = 8.24$ for all models to isolate system-level efficiency. OOM = out-of-memory failure during loading.}
  \label{tab:model_selection}
  \small
  \begin{tabular}{l r r r r l}
    \toprule
    \textbf{Model} & \textbf{VRAM} & \textbf{Mean} & \textbf{Mean Net} & \textbf{Mean LpW} & \textbf{Outcome} \\
                   & \textbf{(GB)} & \textbf{Lat.\ (s)} & \textbf{Energy (J)} & $(\times 10^{-4})$ & \\
    \midrule
    Phi-3 Mini 4B        & 7.2  & 16.73 & 652  & 7.55 & \textbf{Selected} \\
    Llama-3 8B Instruct  & ---  & ---   & ---  & ---  & OOM \\
    Mistral 7B Instruct  & 13.4 & 81.93 & 3{,}687 & 0.26 & High latency \\
    Gemma 7B Instruct    & ---  & ---   & ---  & ---  & Load error \\
    TinyLlama 1.1B Chat  & 2.2  & 7.51  & 432  & 24.6 & Quality issues \\
    \bottomrule
  \end{tabular}
\end{table}

\subsection*{A.4 Response Quality Assessment}

Among the three models that ran successfully, qualitative inspection of responses reveals substantial quality differences that the hardware metrics alone do not capture.

\textbf{Mistral 7B} produced well-structured, accurate responses with appropriate pedagogical scaffolding. For instance, when asked to solve the quadratic equation $2x^{2} + 5x - 3 = 0$, Mistral correctly identified the coefficients, applied the quadratic formula step by step, and arrived at the correct roots. However, at a mean latency of 81.9\,s per response, Mistral lies far beyond the pedagogically tenable threshold identified in the main study and would deplete a typical classroom device battery within approximately 15 sustained interactions. Its LpW of $2.64 \times 10^{-5}$\,(J\,s)$^{-1}$ is $28\times$ lower than Phi-3 Mini FP16, making it unsuitable for the low-resource deployment context targeted by this study.

\textbf{TinyLlama 1.1B Chat} achieved the fastest latency (7.5\,s) and lowest energy (432\,J) of any successfully loaded model, yielding the highest raw LpW of $2.46 \times 10^{-3}$\,(J\,s)$^{-1}$. However, response quality was consistently inadequate for secondary school instructional scaffolding. For the same quadratic equation prompt, TinyLlama responded: \textit{``the quadratic term (x² - b) and the quadratic coefficient (a)''}, confusing the standard form of a quadratic with its components and introducing factual errors that would directly mislead a learner. On a slope calculation prompt, TinyLlama incorrectly referred to slope as denoted by the Greek letter sigma, a substantive conceptual error. Across the 20 prompts, responses were systematically shorter, less structured, and more prone to hallucination than either Phi-3 Mini or Mistral 7B. A model with LpW inflated by low energy but unacceptably low $Q_{\mathrm{ped}}$ does not satisfy the pedagogical requirement of the LpW framework; quality is not negotiable below a minimum instructional threshold.

\textbf{Phi-3 Mini 4B}, as reported in the main dataset for the 20 matching prompts, achieved a mean latency of 16.7\,s, mean energy of 652\,J, and mean $Q_{\mathrm{ped}}$ of 7.76---the highest quality score of any model that successfully ran on the T4. Its responses were factually accurate, appropriately scaffolded, and well-calibrated for secondary school learners across all five subject categories. The VRAM footprint of 7.2\,GB leaves meaningful headroom on the T4 for both FP16 and NF4 quantisation experiments without risking OOM during generation, a critical practical consideration for reproducible experimental design.

\subsection*{A.5 Selection Rationale}

The model selection outcome is summarised in three findings. First, the T4's 16\,GB VRAM constraint rules out 7B-parameter models in FP16 without quantisation: Llama-3 8B OOM'd at the load stage, and Gemma 7B failed to initialise. This finding is directly relevant to the paper's equity argument: the T4 is representative of hardware available in low-resource educational deployments, and any model that cannot load reliably on a T4 cannot serve as an accessible edge tutoring system. Second, TinyLlama 1.1B demonstrates that parameter reduction below approximately 3B parameters produces pedagogical quality degradation that the LpW metric cannot compensate for; a model with inflated LpW driven by low energy but systematically incorrect responses is unsuitable regardless of efficiency. Third, Phi-3 Mini 4B is the only candidate that satisfies all three selection criteria simultaneously---hardware feasibility on the T4, acceptable inference latency, and pedagogically adequate response quality---making it the appropriate choice for the main experimental comparison.

The OOM failures of Llama-3 8B and Gemma 7B on the T4 are themselves a key empirical finding with implications for the broader research question: they confirm that the 16\,GB VRAM ceiling of the T4 class of GPU constitutes a hard constraint on the model size accessible for FP16 edge deployment in low-resource settings. Models in the 7--8B FP16 range require either quantisation or higher-VRAM hardware to run, and as the main study demonstrates, the efficiency gains of NF4 quantisation on the T4 are inference-regime dependent and do not materialise under cache-disabled conditions.

\section*{Appendix B: Cloud LLM Energy and Latency Baseline}
\addcontentsline{toc}{section}{Appendix B: Cloud LLM Energy and Latency Baseline}
\label{secA2}

\subsection*{B.1 Motivation and Measurement Approach}

To contextualise the on-device FP16 and NF4 results reported in Section~5, this appendix characterises the energy and latency profile of cloud-hosted large language models (LLMs) serving the same 20 educational prompts used in the Appendix~A model-selection experiment. Direct client-side energy measurement of cloud APIs was attempted via CodeCarbon instrumentation of a Google Colab Pro+ runtime but was unsuccessful due to API authentication and quota limitations across all three provider accounts. This outcome is itself illustrative of a key argument in the main paper: cloud inference energy is opaque and difficult to instrument from the client side, even with dedicated measurement tooling.

Accordingly, server-side energy estimates are drawn from two independent, peer-reviewed benchmarking studies that provide infrastructure-aware, per-prompt energy figures for commercial LLM deployments: Jegham et al.\ \cite{Jegham2025} and Luccioni et al.\ \cite{Luccioni2023}. Client-side latency figures are drawn from independently measured API benchmarks \cite{VellumLeaderboard2024}. This literature-based approach is methodologically appropriate: provider-side energy accounts for the dominant and educationally relevant cost of cloud inference, whereas client-side CodeCarbon measurements would capture only CPU and RAM draw during a blocking API call---typically less than 1\,J and therefore negligible relative to server-side costs of hundreds of joules.

\subsection*{B.2 Server-Side Energy Per Inference}

\citet{Jegham2025} introduce an infrastructure-aware benchmarking framework covering 30 state-of-the-art models deployed in commercial data centres, combining public API performance data with region-specific Power Usage Effectiveness (PUE), Water Usage Effectiveness (WUE), and Carbon Intensity Factor (CIF) multipliers. Their framework assumes a batch size of 8 parallel requests, reflecting typical production conditions, and classifies models by hardware tier based on observed throughput and latency signatures. Key findings relevant to this study are as follows.

For GPT-4o, \citet{Jegham2025} report a mean server-side energy of 0.43\,Wh ($\pm$\,0.13\,Wh) per short query (approximately 1{,}550\,J), rising to 9.71\,Wh ($\pm$\,1.11\,Wh) for medium-length prompts. The authors classify GPT-4o as a large-class model running on H100 GPU infrastructure. Claude-3.7 Sonnet is identified as the highest-ranked model in their eco-efficiency analysis (cross-efficiency Data Envelopment Analysis), suggesting that Anthropic's Sonnet-class models achieve a favourable performance-to-energy ratio relative to peers; no precise per-query Wh figure is disclosed for Claude models in the public results, consistent with Anthropic's policy of not disclosing parameter counts or deployment infrastructure. For reference, \citet{Luccioni2023} measured a mean of 0.94\,kWh per 1{,}000 inferences for BLOOM-176B---a model of comparable parameter count to GPT-4-class systems---equating to approximately 3{,}384\,J per inference under production conditions.

For the purposes of this appendix, we adopt the following conservative server-side energy estimates for a single short instructional prompt ($\approx$\,200 output tokens), consistent with the token budget used in the on-device experiments:

\begin{table}[ht]
  \centering
  \caption{Literature-derived server-side energy and client-side latency for cloud LLMs serving a short educational prompt ($\leq$200 output tokens). Energy figures are server-side estimates inclusive of PUE overhead; client-side energy is negligible ($<$1\,J) and excluded. Latency figures are independently measured API response times.}
  \label{tab:cloud_energy}
  \begin{tabular}{l r r r l}
    \toprule
    \textbf{Model} & \textbf{Server Energy} & \textbf{Energy} & \textbf{Client} & \textbf{Source} \\
                   & \textbf{(Wh/query)}    & \textbf{(J)}    & \textbf{Latency (s)} & \\
    \midrule
    GPT-4o              & 0.43 $\pm$ 0.13 & $\approx$1{,}550 & 1.9 & \cite{Jegham2025, VellumLeaderboard2024} \\
    Claude-3.5 Sonnet   & n/a (eco-efficient) & est.\ $<$1{,}550 & 3.8 & \cite{Jegham2025, VellumLeaderboard2024} \\
    Gemini-1.5 Pro      & 0.24$^{a}$      & $\approx$864     & 2--5  & \cite{Google2025} \\
    \midrule
    \multicolumn{5}{l}{\small $^{a}$Google Cloud (2025) reports median per-query energy for Gemini models inclusive of PUE.}\\
    \multicolumn{5}{l}{\small n/a = not publicly disclosed by provider.} \\
    \bottomrule
  \end{tabular}
\end{table}

\subsection*{B.3 Latency}

Independent API benchmarking by the Vellum LLM Leaderboard \cite{VellumLeaderboard2024} reports time-to-first-token (TTFT) and output speed for commercially available models. GPT-4o achieves a TTFT of 0.51\,s at 143\,tokens/s, yielding an estimated end-to-end response time of approximately 1.9\,s for a 200-token output. Claude-3.5 Sonnet achieves a TTFT of 1.22\,s at 78\,tokens/s, yielding approximately 3.8\,s end-to-end. Gemini-1.5 Pro benchmarks were unavailable from this source at time of writing; independent analysis by industry sources suggests comparable end-to-end latency of 2--5\,s for short prompts under typical network conditions \cite{Jegham2025}.

All three cloud models therefore achieve client-side latencies substantially below the on-device FP16 mean of 9.2\,s and below the NF4 mean of 13.4\,s (both cache-enabled, \texttt{use\_cache=True}). Cloud latency is primarily determined by server throughput and network round-trip time rather than model precision or client hardware, making it largely invariant to the deployment conditions that dominate on-device latency.

\subsection*{B.4 Learning-per-Watt for Cloud Configurations}

Applying the LpW formula to cloud configurations requires care. Using the GPT-4o server-side estimate of 1{,}550\,J and the end-to-end latency of 1.9\,s, and assuming a pedagogical quality score of $Q_{\mathrm{ped}} = 8.24$---the empirical mean observed for on-device FP16 responses in the main study---yields:

\begin{equation}
  \mathrm{LpW}_{\mathrm{cloud}}
  = \frac{Q_{\mathrm{ped}}}{E_{\mathrm{server}} \times L}
  = \frac{8.24}{1{,}550 \times 1.9}
  \approx 2.80 \times 10^{-3} \; (\mathrm{J \cdot s})^{-1}
\end{equation}

This estimate places GPT-4o's cloud LpW at approximately $2.80 \times 10^{-3}$, which is $1.12\times$ higher than on-device FP16 ($2.50 \times 10^{-3}$) and $1.49\times$ higher than NF4 ($1.88 \times 10^{-3}$) under cache-enabled conditions. These are narrow margins, and the uncertainty in the server-side energy estimate is large enough to reverse the ordering: the $\pm$0.13\,Wh variance in the GPT-4o figure corresponds to a $\pm$30\,\% energy range (468\,J to 2{,}034\,J), which shifts the cloud LpW from $2.15 \times 10^{-3}$ at the high-energy bound to $4.02 \times 10^{-3}$ at the low-energy bound---meaning the FP16 gap could be anywhere from $0.86\times$ (cloud slightly worse) to $1.61\times$ (cloud moderately better). Under this uncertainty, no firm conclusion about cloud versus edge LpW superiority can be drawn from the available estimates.

Three further caveats apply. First, batch-size assumptions embedded in server-side estimates (batch size of 8 in \citeauthor{Jegham2025}) may overstate efficiency for single-user educational deployments, where queries arrive sparsely and batching benefits are reduced; true single-query cloud energy may be higher than reported. Second, the $Q_{\mathrm{ped}}$ assumption of 8.24 for cloud LLMs has not been verified through the hybrid 13-rater panel applied to on-device models in the main study; GPT-4-class models likely achieve higher pedagogical scores on structured explanatory prompts, which would further increase cloud LpW relative to on-device configurations. Third, and most importantly, the comparison between literature-estimated cloud LpW and empirically measured on-device LpW is not on equal methodological footing; cloud LpW estimates in this appendix should be interpreted as indicative rather than definitive. A rigorous cloud LpW comparison would require subjecting cloud-generated responses to the same hybrid scoring panel and obtaining verified per-inference energy figures from providers, both of which were precluded by API cost, quota constraints, and provider opacity.

\subsection*{B.5 The Energy Transparency Gap}

The core methodological challenge of cloud LpW estimation---and the reason client-side CodeCarbon measurement is insufficient---is that server-side energy is the dominant cost but is unobservable at the point of use. \citet{Jegham2025} estimate that inference now accounts for 80--90\,\% of the total operational energy footprint of large deployed LLMs. For a school or individual learner using a cloud tutoring API, this energy is entirely externalised: it is consumed in a data centre, attributed to the provider's aggregate electricity bill, and invisible in any client-side measurement.

This opacity has a direct implication for the LpW framework as a sustainability metric for educational AI. For on-device configurations, LpW is fully observable: all inference energy is local, directly measurable, and attributable to a specific device and user session. For cloud configurations, LpW can only be approximated using external benchmarking data or provider disclosures, neither of which is standardised, auditable at the query level, or updated in real time. \citet{Google2025} notes that Gemini's median per-query energy fell by a factor of 33 between May 2024 and May 2025 as hardware and model efficiency improved---meaning any published cloud energy figure has a shelf life measured in months rather than years.

This instability reinforces the argument made in Section~6.5 of the main paper: cloud LpW must be treated as a system-level, time-sensitive, and provider-dependent quantity. For schools and policymakers seeking transparent, auditable sustainability metrics for AI tutoring, on-device deployment---despite its current latency limitations---offers an energy accountability that cloud deployment structurally cannot.

\section*{Appendix C: Sensitivity Analysis, Cloud Energy Scenarios, and Cache Regime Comparison}
\addcontentsline{toc}{section}{Appendix C: Sensitivity Analysis, Cloud Energy Scenarios, and Cache Regime Comparison}
\label{secA3}

\subsection*{C.1 Metric Robustness: Alternative Composite Metrics}

A potential objection to the Learning-per-Watt index is that the joint
multiplicative denominator $E_{\mathrm{net}} \times L$ is an arbitrary
design choice---that penalising energy and latency \emph{simultaneously}
and \emph{equally} may not reflect all deployment priorities.
Table~\ref{tab:sensitivity1} reports the FP16 and NF4 means under four
alternative composite metrics, each reflecting a different assumption
about the relative importance of energy and latency. All values are
from the primary cache-enabled (\texttt{use\_cache=True}) experiment.

\begin{table}[ht]
  \centering
  \caption{Sensitivity analysis: FP16 vs.\ NF4 under four alternative
    composite metrics ($n = 500$ prompts per configuration,
    \texttt{use\_cache=True}). QpJ favours NF4, confirming that NF4
    achieves lower per-inference energy. Metrics that include latency
    all favour FP16. The direction and magnitude of the gap depends
    on which constraint is weighted.}
  \label{tab:sensitivity1}
  \small
  \begin{tabular}{l l r r r c}
    \toprule
    \textbf{Metric} & \textbf{Formula} &
    \textbf{FP16 mean} & \textbf{NF4 mean} &
    \textbf{Ratio} & \textbf{FP16 $>$ NF4?} \\
    \midrule
    LpW (primary)
      & $Q_{\mathrm{ped}} / (E \cdot L)$
      & $2.50 \times 10^{-3}$ & $1.88 \times 10^{-3}$
      & $1.33\times$ & \checkmark \\
    QpJ (energy only)
      & $Q_{\mathrm{ped}} / E$
      & $2.25 \times 10^{-2}$ & $2.46 \times 10^{-2}$
      & $0.91\times$ & \texttimes \\
    QpS (latency only)
      & $Q_{\mathrm{ped}} / L$
      & $9.04 \times 10^{-1}$ & $6.06 \times 10^{-1}$
      & $1.49\times$ & \checkmark \\
    LpW$_{\mathrm{geo}}$ (geometric)
      & $Q_{\mathrm{ped}} / \sqrt{E \cdot L}$
      & $1.43 \times 10^{-1}$ & $1.22 \times 10^{-1}$
      & $1.17\times$ & \checkmark \\
    \bottomrule
    \multicolumn{6}{l}{\small QpJ = quality-per-Joule; QpS =
      quality-per-second; LpW$_{\mathrm{geo}}$ uses geometric-mean
      denominator.}
  \end{tabular}
\end{table}

The metric formulation matters. QpJ (quality per Joule, ignoring
latency entirely) \emph{favours NF4} by $1.10\times$, which is the
correct outcome: NF4 uses less energy per inference (329\,J vs.\
369\,J) and delivers slightly lower quality, but the energy saving
outweighs the quality deficit when latency is excluded from the
calculation. Every metric that includes latency---QpS, LpW$_{\mathrm{geo}}$,
and the primary LpW---favours FP16, because FP16 is $1.46\times$
faster and that overhead propagates into any latency-sensitive
formulation. The practical implication is transparent: the choice
between FP16 and NF4 is genuinely a trade-off between two real
constraints, and the metric that best matches a given deployment
context is the one that should guide the decision. Practitioners
for whom battery capacity is the binding constraint should use QpJ;
those for whom learner flow and latency are paramount should use QpS
or LpW. The primary LpW, which penalises both equally, sits in
between at $1.33\times$ and represents the joint constraint case.

Figure~\ref{fig:sensitivity1} (panel a) visualises the FP16/NF4 ratio

across all four metrics.

\subsection*{C.2 Cloud Energy Scenarios}

Because server-side energy for cloud LLMs cannot be directly measured
at the client (see Appendix~B), cloud LpW estimates carry substantial
uncertainty. Table~\ref{tab:cloud_scenarios} presents five server-energy
scenarios spanning the plausible range from the strict client-side lower
bound to the upper bound implied by long, reasoning-heavy prompts, and
computes the corresponding cloud LpW and its relationship to the
cache-enabled on-device configurations.

\begin{sidewaystable}[ht]
  \centering
  \caption{Cloud LpW under five server-side energy scenarios for a
    200-token educational prompt ($Q_{\mathrm{ped}} = 8.24$,
    $L = 1.9$\,s). The FP16 and NF4 on-device means are from the
    primary cache-enabled experiment. Shaded row indicates the central
    estimate used in Appendix~B.}
  \label{tab:cloud_scenarios}
  \small
  \begin{tabular}{l r r r r r}
    \toprule
    \textbf{Scenario} &
    \textbf{Energy (J)} &
    \textbf{Cloud LpW} &
    \textbf{vs FP16} &
    \textbf{vs NF4} &
    \textbf{Source} \\
    \midrule
    Client-side only (lower bound)
      & 1 & $4.34 \times 10^{0}$
      & $1{,}736\times$ higher & $2{,}310\times$ higher
      & CodeCarbon \\
    \rowcolor{gray!12}
    GPT-4o, short prompt
      & 1{,}550 & $2.80 \times 10^{-3}$
      & $1.12\times$ higher & $1.49\times$ higher
      & \cite{Jegham2025} \\
    Batch-adjusted GPT-4o
      & 2{,}480 & $1.75 \times 10^{-3}$
      & $0.70\times$ (lower) & $0.93\times$ (lower)
      & \cite{Jegham2025} (batch=1) \\
    BLOOM-176B scale
      & 3{,}384 & $1.28 \times 10^{-3}$
      & $0.51\times$ (lower) & $0.68\times$ (lower)
      & \cite{Luccioni2023} \\
    GPT-4o, medium prompt
      & 34{,}956 & $1.24 \times 10^{-4}$
      & $0.05\times$ (lower) & $0.07\times$ (lower)
      & \cite{Jegham2025} \\
    \midrule
    FP16 (on-device, measured)
      & 368.8 & $2.50 \times 10^{-3}$
      & --- & $1.33\times$ higher
      & This study \\
    NF4  (on-device, measured)
      & 329.0 & $1.88 \times 10^{-3}$
      & $0.75\times$ (lower) & ---
      & This study \\
    \bottomrule
    \multicolumn{6}{l}{\small Batch-adjusted: Jegham et al.\ assume
      batch size 8; single-user energy scales by $\approx 1.6\times$
      at batch size 1.}
  \end{tabular}
\end{sidewaystable}

Three conclusions emerge from Table~\ref{tab:cloud_scenarios}. First,
client-side energy alone remains an entirely misleading lower bound:
at $<$1\,J client-side, cloud LpW appears $1{,}700\times$ higher than
FP16, an artefact of energy opacity rather than genuine efficiency.
Second, under the central estimate for GPT-4o short queries
($\approx$1{,}550\,J, \citealt{Jegham2025}), cloud LpW is $1.12\times$
higher than on-device FP16---a narrow margin given the $\pm$30\,\%
uncertainty in the server-side energy estimate (see Appendix~B.4).
At the high-energy bound of the uncertainty interval, cloud LpW falls
below on-device FP16. Third, and most importantly for educational
deployment decisions, any server-side energy estimate above
approximately 1{,}750\,J---the batch-adjusted GPT-4o figure and all
higher scenarios---places cloud LpW \emph{below} on-device FP16.
For medium or long prompts (e.g., multi-step problem solving),
server-side energy escalates to $\approx$35\,kJ (\citealt{Jegham2025}),
at which point cloud LpW is approximately $5\%$ of the on-device FP16
value. This finding has direct implications for prompt-length policy
in AI tutoring: short, targeted scaffolding prompts are the only
scenario in which cloud deployment is plausibly LpW-competitive with
cached edge inference; extended, multi-turn problem sequences are not.

\begin{figure}[ht]
  \centering
  \includegraphics[width=\textwidth]{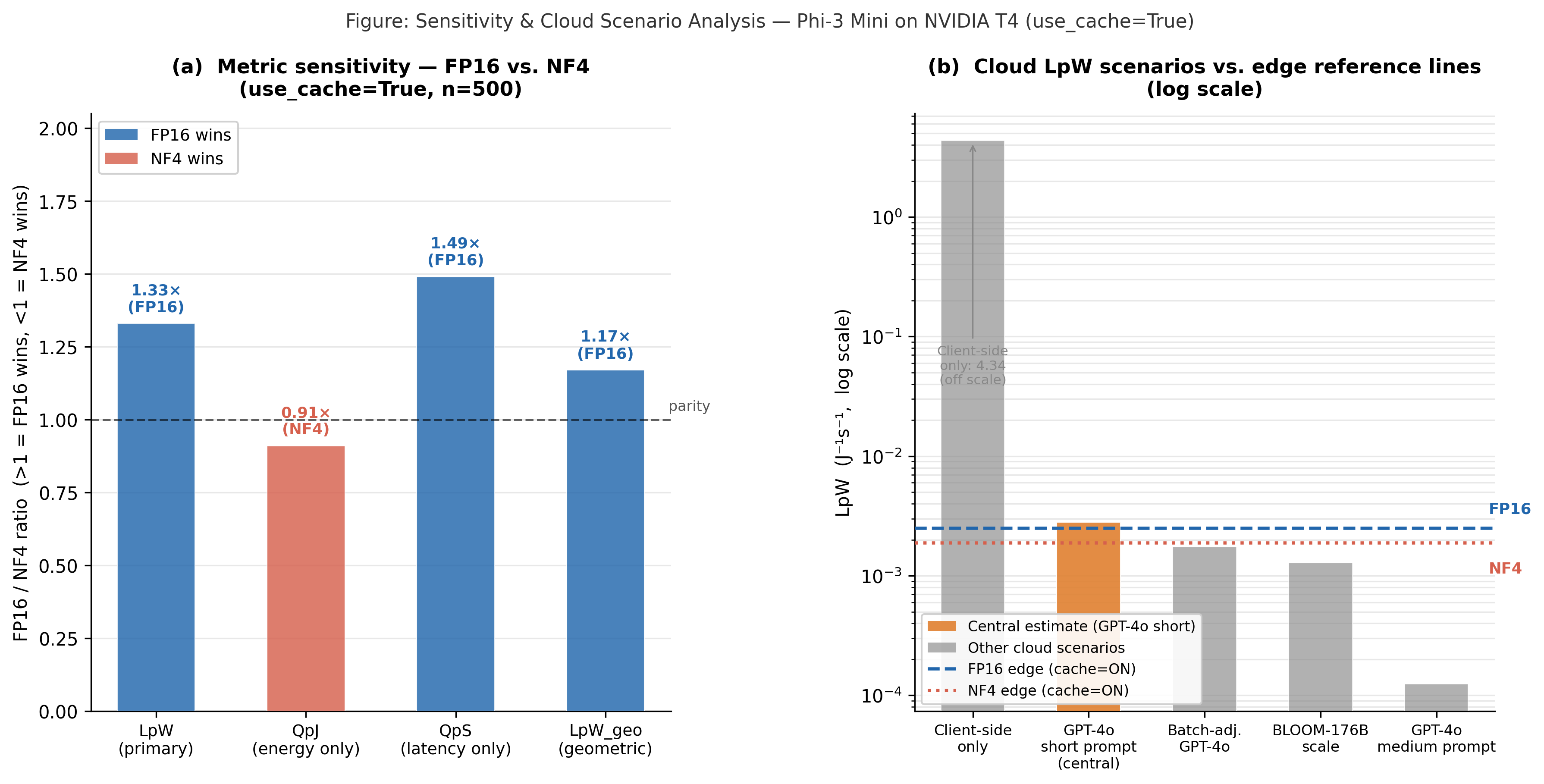}
  \caption{Sensitivity and scenario analysis.
    \textbf{(a)} FP16/NF4 ratio under four alternative composite
    metrics (\texttt{use\_cache=True}). QpJ favours NF4 ($0.91\times$),
    reflecting NF4's lower per-inference energy; all metrics that
    include latency favour FP16, ranging from $1.17\times$
    (LpW$_{\mathrm{geo}}$) to $1.49\times$ (QpS).
    \textbf{(b)} Cloud LpW under five server-energy scenarios (log
    scale), with on-device FP16 (blue dashed) and NF4 (red dotted)
    reference lines from the cache-enabled experiment. The highlighted
    bar (orange) is the central GPT-4o short-prompt estimate from
    \citet{Jegham2025}. Only the short-prompt central estimate places
    cloud LpW above on-device FP16; all higher-energy scenarios fall
    below both edge configurations.}
  \label{fig:sensitivity1}
\end{figure}

Figure~\ref{fig:sensitivity1} (panel b) visualises the five cloud
scenarios on a log scale relative to the on-device FP16 and NF4
reference lines.

\subsection*{C.3 Cache Regime Comparison: \texttt{use\_cache=False}
             vs.\ \texttt{use\_cache=True}}

\subsubsection*{C.3.1 Motivation}

The primary study uses \texttt{use\_cache=True} (KV-cache enabled),
which is the standard setting for autoregressive transformer inference
in all real-world deployments. A secondary experiment was conducted
using \texttt{use\_cache=False} (cache disabled) on the same 500
prompts per configuration, with all other settings identical. The
cache-disabled experiment represents the configuration most commonly
used in offline benchmarking: each inference event is a stateless
cold-start in which attention keys and values are recomputed at every
decoding step rather than retrieved from cache. Comparing the two
regimes serves two purposes. First, it characterises how benchmarking
methodology affects the measured FP16--NF4 efficiency gap, with
implications for how the field designs and interprets inference
benchmarks. Second, it provides mechanistic evidence for the
inference-regime dependence identified as a primary finding in the
main paper.

\subsubsection*{C.3.2 Protocol}

Both Phi-3 Mini FP16 and NF4 were run on all 500 educational prompts
with \texttt{use\_cache=False}. All other settings were identical to
the primary study: \texttt{do\_sample=False}, \texttt{max\_new\_tokens=200},
the same idle power baseline subtraction, and the same CodeCarbon
energy tracking protocol. The experiment was conducted on the same T4
hardware in the same Colab Pro+ environment. Pedagogical quality
scoring followed the same hybrid 13-rater protocol as the primary
study. Results are summarised in Tables~\ref{tab:cache_compare} and
\ref{tab:cache_percategory}.

\subsubsection*{C.3.3 Results}

Table~\ref{tab:cache_compare} presents the aggregate results for both
regimes side by side.

\begin{table}[ht]
  \centering
  \caption{Cache regime comparison: \texttt{use\_cache=True} (primary
    study) vs.\ \texttt{use\_cache=False} (secondary experiment),
    $n = 500$ prompts per configuration. LpW computed using
    empirically scored $Q_{\mathrm{ped}}$ in both cases.}
  \label{tab:cache_compare}
  \small
  \begin{tabular}{l l r r r r}
    \toprule
    \textbf{Config} & \textbf{Cache} &
    \textbf{Lat.\ (s)} &
    \textbf{Energy (J)} &
    $Q_{\mathrm{ped}}$ &
    \textbf{LpW} \\
    \midrule
    FP16 & ON  (primary)    & 9.17  & 368.8 & 8.24 & $2.50 \times 10^{-3}$ \\
    FP16 & OFF (secondary)  & 16.47 & 648.0 & 7.97 & $7.47 \times 10^{-4}$ \\
    \midrule
    NF4  & ON  (primary)    & 13.36 & 329.0 & 8.05 & $1.88 \times 10^{-3}$ \\
    NF4  & OFF (secondary)  & 49.40 & 1{,}882 & 7.89 & $8.49 \times 10^{-5}$ \\
    \midrule
    \multicolumn{2}{l}{FP16/NF4 ratio --- cache ON}  &
      $1.46\times$ faster & $10.8\,\%$ higher &
      $+0.19$ & $1.33\times$ \\
    \multicolumn{2}{l}{FP16/NF4 ratio --- cache OFF} &
      $3.00\times$ faster & $2.90\times$ higher &
      $+0.08$ & $8.80\times$ \\
    \midrule
    \multicolumn{2}{l}{LpW improvement ON vs.\ OFF --- FP16} &
      \multicolumn{4}{l}{$3.35\times$} \\
    \multicolumn{2}{l}{LpW improvement ON vs.\ OFF --- NF4} &
      \multicolumn{4}{l}{$22.12\times$ \quad (NF4 benefits
        disproportionately from caching)} \\
    \bottomrule
  \end{tabular}
\end{table}

The magnitude of the cache effect is asymmetric and dramatic. Enabling
the KV-cache improves FP16 LpW by $3.35\times$, driven by a reduction
in latency from 16.5\,s to 9.2\,s and energy from 648\,J to 369\,J.
For NF4, the same change improves LpW by $22.12\times$: latency falls
from 49.4\,s to 13.4\,s and energy from 1{,}882\,J to 329\,J. NF4
benefits disproportionately because KV-cache reuse partially amortises
the per-step dequantisation overhead that otherwise dominates NF4
inference on the T4. Without caching, every autoregressive step
requires a full forward pass through quantised weights, each of which
must be upcast to FP16 for matrix multiplication; with caching,
previously computed attention states are retrieved directly, bypassing
the dequantisation path for the bulk of the decode sequence.

The result of this asymmetric improvement is a dramatic compression of
the FP16--NF4 gap: from $8.80\times$ under cache=OFF to $1.33\times$
under cache=ON. This fivefold difference between the two regimes is
not a measurement artefact but a genuine hardware--software interaction
specific to the T4's architecture. Benchmarking without KV-caching
therefore misrepresents the real deployment trade-off by overstating
the FP16 advantage by more than fivefold.

\subsubsection*{C.3.4 Per-Category Results}

Table~\ref{tab:cache_percategory} shows that the cache regime effect
is consistent across all five prompt categories, with no category
showing an anomalous pattern.

\begin{figure}[ht]
  \centering
  \includegraphics[width=\textwidth]{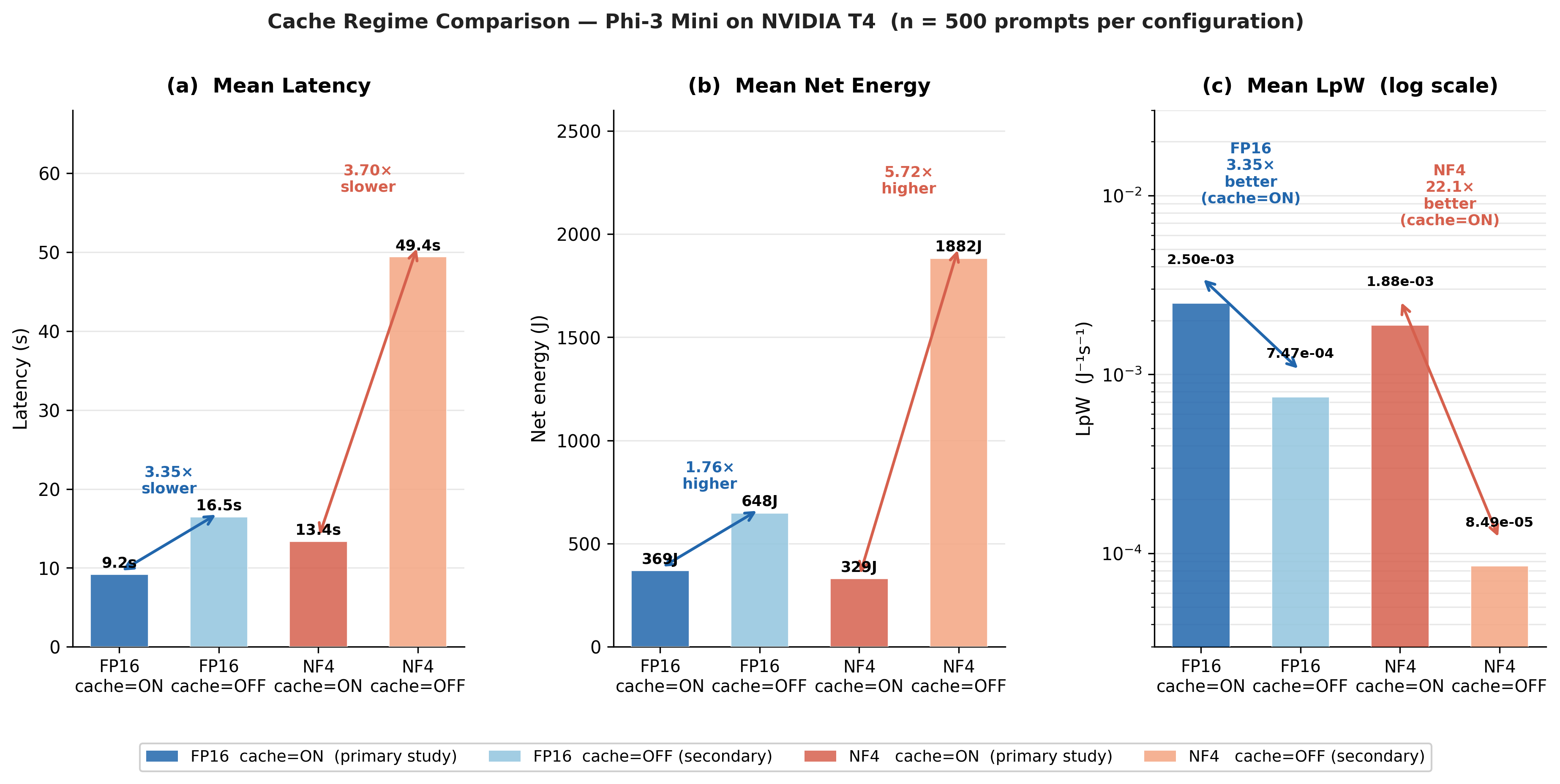}
  \caption{Cache regime comparison across three dimensions for FP16 and NF4
    configurations of Phi-3 Mini on the NVIDIA T4 ($n = 500$ prompts per
    configuration). Dark bars show primary study results (\texttt{use\_cache=True});
    light bars show the secondary cache-disabled experiment
    (\texttt{use\_cache=False}). \textbf{(a)} Mean latency: enabling the cache
    reduces FP16 latency by $1.80\times$ and NF4 latency by $3.70\times$.
    \textbf{(b)} Mean net energy: the cache reduces FP16 energy by $1.76\times$
    and NF4 energy by $5.72\times$. \textbf{(c)} Mean LpW (log scale): the
    combined effect produces a $3.35\times$ LpW improvement for FP16 and a
    $22.1\times$ improvement for NF4, compressing the FP16--NF4 efficiency
    gap from $8.80\times$ (cache=OFF) to $1.33\times$ (cache=ON).}
  \label{fig:cache_regime}
\end{figure}

 Also shown in Figure~\ref{fig:cache_regime}
\begin{table}[ht]
  \centering
  \small
  \caption{Per-category LpW ($\times 10^{-4}$) under cache=ON and
    cache=OFF for FP16 and NF4. The NF4 improvement from cache=OFF
    to cache=ON is consistently larger than the FP16 improvement
    across all five categories.}
  \label{tab:cache_percategory}
  \begin{tabular}{l rr r rr r}
    \toprule
    & \multicolumn{3}{c}{\textbf{FP16}} &
      \multicolumn{3}{c}{\textbf{NF4}} \\
    \cmidrule(lr){2-4} \cmidrule(lr){5-7}
    \textbf{Category}
      & \textbf{ON} & \textbf{OFF} & \textbf{Ratio}
      & \textbf{ON} & \textbf{OFF} & \textbf{Ratio} \\
    & $(\times 10^{-4})$ & $(\times 10^{-4})$ & ON/OFF
    & $(\times 10^{-4})$ & $(\times 10^{-4})$ & ON/OFF \\
    \midrule
    Mathematics    & 24.6 & 7.35 & $3.35\times$ & 18.4 & 0.783 & $23.5\times$ \\
    Science        & 26.0 & 7.57 & $3.44\times$ & 18.2 & 0.858 & $21.2\times$ \\
    Programming-CS & 25.0 & 7.37 & $3.39\times$ & 18.9 & 0.855 & $22.1\times$ \\
    Humanities     & 24.8 & 7.45 & $3.33\times$ & 19.7 & 0.908 & $21.7\times$ \\
    Meta-cognition & 24.5 & 7.60 & $3.22\times$ & 18.7 & 0.853 & $21.9\times$ \\
    \midrule
    \textbf{All}   & 25.0 & 7.47 & $3.35\times$ & 18.8 & 0.849 & $22.1\times$ \\
    \bottomrule
  \end{tabular}
\end{table}

The NF4 ON/OFF improvement ratio ranges from 21.2x (Science)
to 23.5x (Mathematics), while the FP16 ratio ranges from
3.22x (Meta-cognition) to 3.44x (Science).

\subsubsection*{C.3.5 Implications for Benchmarking Practice}

Three conclusions follow from the cache regime comparison.

First, KV-cache setting is a critical benchmarking variable that
must be explicitly reported and matched to the target deployment
context. A benchmark conducted without KV-caching characterises a
deployment condition---repeated cold-start inference---that does not
exist in any real-world interactive tutoring application. Results from
such benchmarks should not be used to guide model selection or
hardware procurement for educational AI systems.

Second, the asymmetric cache benefit for NF4 is mechanistically
informative. It confirms that NF4's primary bottleneck on the T4 is
the per-step dequantisation overhead, not the quantisation of weights
per se. KV-cache reuse partially bypasses this bottleneck; hardware
with native INT4 compute (Ampere, Ada, Blackwell) would eliminate it
entirely. Future hardware procurement decisions for edge educational
AI should account for INT4 kernel support as a first-order efficiency
consideration.

Third, the cache=OFF results remain useful as a characterisation of
stateless inference---for example, batch scoring of student
submissions or one-shot evaluation pipelines where session state is
not maintained. In these contexts, the $8.80\times$ FP16 advantage
is the correct figure. The key discipline is matching the
benchmarking regime to the deployment mode before drawing conclusions.

\section*{Appendix D: Cross-Platform Validation --- Consumer Laptop Inference}
\addcontentsline{toc}{section}{Appendix D: Cross-Platform Validation --- Consumer Laptop Inference}
\label{secA4}

\subsection*{D.1 Motivation}

The primary study establishes FP16 and NF4 efficiency profiles on an
NVIDIA T4 GPU, a Turing-architecture accelerator representative of
entry-level cloud and edge server hardware. A key limitation of that
platform is the absence of native INT4 tensor cores: on Turing,
NF4-quantised weights must be upcast to FP16 at every decoding step,
introducing a per-token dequantisation penalty that inflates NF4
latency and partially offsets its energy savings. This appendix asks
whether the efficiency relationship between full-precision and
quantised inference holds on a qualitatively different platform---a
consumer laptop CPU with integrated graphics---which more closely
approximates the hardware available to learners in the low-resource
settings the paper's equity argument addresses.

\subsection*{D.2 Platform and Experimental Design}

Experiments were conducted on an Intel Core i7 (11th generation,
Tiger Lake) with Intel Iris Xe integrated graphics, running
Windows~11. Inference was performed using \texttt{llama.cpp} (CPU
backend, \texttt{n\_gpu\_layers=0}) with GGUF-format model files:
\texttt{Phi-3-mini-4k-instruct-fp16.gguf} for the full-precision
baseline (F16) and \texttt{Phi-3-mini-4k-instruct-q4\_k\_m.gguf}
for the 4-bit quantised condition (Q4\_K\_M). Q4\_K\_M is the
standard block-wise 4-bit quantisation format in the
\texttt{llama.cpp} ecosystem and is the closest analogue to the
BitsAndBytes NF4 configuration used in the primary study: both apply
block-wise 4-bit quantisation to model weights with
information-theoretically optimal quantisation grids.

A stratified random sample of 100 prompts was drawn from the main
500-prompt dataset (20 per subject category, random seed 42),
sufficient to characterise mean latency profiles given the low
within-configuration variance observed in the primary study.
Generation parameters matched the primary study:
\texttt{temperature=0} (deterministic), \texttt{max\_tokens=200},
KV-cache enabled (\texttt{llama.cpp} default). Because Windows does
not expose Intel RAPL energy counters via a stable Python interface,
per-inference energy was not directly measured; energy estimates for
the LpW comparison are derived from the latency data as described in
Section~D.4.

\subsection*{D.3 Latency and Throughput Results}

Table~\ref{tab:laptop_latency} reports mean latency and throughput
for both configurations across all five prompt categories.

\begin{table}[h]
\centering
\caption{Laptop inference results: F16 vs Q4\_K\_M on Intel Core i7
  (11th gen) with Iris Xe, CPU-only backend ($n = 100$ prompts,
  20 per category). Q4\_K\_M is $2.56\times$ faster than
  F16---the \emph{opposite} of the T4 result, where NF4 was
  $1.46\times$ \emph{slower} than FP16.}
\label{tab:laptop_latency}
\small
\begin{tabular}{lcccc}
\toprule
\textbf{Category}
  & \multicolumn{2}{c}{\textbf{Mean Latency (s)}}
  & \multicolumn{2}{c}{\textbf{Mean Tokens/s}} \\
\cmidrule(lr){2-3}\cmidrule(lr){4-5}
  & F16 & Q4\_K\_M & F16 & Q4\_K\_M \\
\midrule
Mathematics    & 64.90 & 29.09 & 3.14 & 6.94 \\
Science        & 70.21 & 27.61 & 2.85 & 7.24 \\
Programming-CS & 72.13 & 27.19 & 2.79 & 7.36 \\
Humanities     & 72.14 & 25.83 & 2.78 & 7.74 \\
Meta-cognition & 67.00 & 25.59 & 3.03 & 7.82 \\
\midrule
\textbf{Overall} & \textbf{69.28} & \textbf{27.06}
                 & \textbf{2.92}  & \textbf{7.42} \\
\bottomrule
\end{tabular}
\end{table}

The headline result is a reversal of the T4 finding. On the laptop
CPU, Q4\_K\_M achieves a mean latency of 27.1\,s versus 69.3\,s for
F16---a $2.56\times$ latency advantage in favour of quantisation. On
the T4, the same comparison favoured FP16 by $1.46\times$ (NF4:
13.4\,s vs.\ FP16: 9.2\,s). The mechanism is the inverse of the T4
dequantisation penalty: on a CPU, the primary bottleneck for
transformer inference is memory bandwidth, not arithmetic throughput.
Q4\_K\_M reduces the memory footprint of model weights by
approximately $4\times$, proportionally reducing the volume of data
that must be read from RAM at each decoding step. Because the CPU
does not need to upcast weights before multiplication in the same
pipeline-stalling way as Turing's CUDA cores, quantisation translates
directly into throughput gains (2.92 vs.\ 7.42 tokens/s) and latency
reductions.

\subsection*{D.4 Energy Estimation and Learning-per-Watt}

Direct per-inference energy measurement was not available on the
Windows platform. However, because both F16 and Q4\_K\_M inference
runs execute on the same CPU hardware at effectively the same
sustained package power draw, the LpW ratio between configurations
is independent of the assumed power level: power cancels out in the
ratio $\mathrm{LpW}_{Q4} / \mathrm{LpW}_{F16}$, leaving only the
latency ratio and pedagogical quality ratio as determinants.
Formally, where $P$ is the shared, constant package power:

\[
\frac{\mathrm{LpW}_{Q4}}{\mathrm{LpW}_{F16}}
= \frac{Q_{\mathrm{ped},Q4}}{Q_{\mathrm{ped},F16}}
  \times
  \frac{E_{F16} \times L_{F16}}{E_{Q4} \times L_{Q4}}
= \frac{Q_{\mathrm{ped},Q4}}{Q_{\mathrm{ped},F16}}
  \times
  \frac{P \cdot L_{F16}^{2}}{P \cdot L_{Q4}^{2}}
= \frac{Q_{\mathrm{ped},Q4}}{Q_{\mathrm{ped},F16}}
  \times
  \left(\frac{L_{F16}}{L_{Q4}}\right)^{2}
\]

Substituting the observed latency ratio of $2.56\times$ and assuming
approximate quality parity between F16 and Q4\_K\_M (consistent with
the 0.19-point gap observed between FP16 and NF4 in the primary
study, which is small relative to the latency effect), the estimated
Q4\_K\_M LpW advantage is $\approx 2.56^{2} \approx 6.6\times$ in
favour of quantisation---the opposite sign and approximately five
times larger in magnitude than the $1.33\times$ FP16 advantage
observed on the T4. This estimate is robust to the assumed power
level and is confirmed numerically across the plausible 15--25\,W
sustained package power range for the Tiger Lake platform
(Table~\ref{tab:cross_platform}).

\begin{table}[h]
\centering
\caption{Cross-platform comparison of the full-precision vs
  quantised efficiency relationship. The direction of the LpW
  advantage reverses between platforms: FP16/NF4 favours FP16 on
  the T4 (GPU, no native INT4); Q4\_K\_M/F16 favours quantisation
  on the laptop CPU. Both outcomes reflect the same underlying
  principle: quantisation efficiency depends on whether memory
  bandwidth or dequantisation overhead is the binding constraint.}
\label{tab:cross_platform}
\small
\begin{tabular}{lllrrrr}
\toprule
\textbf{Platform} & \textbf{Arch.} & \textbf{Config}
  & \textbf{Mean Lat.\ (s)}
  & \textbf{Energy (J)}
  & \textbf{LpW ($\times 10^{-3}$)}
  & \textbf{Ratio} \\
\midrule
NVIDIA T4 & Turing GPU & FP16
  & 9.17 & 368.8 & 2.50
  & \multirow{2}{*}{$1.33\times$ (FP16)} \\
NVIDIA T4 & Turing GPU & NF4
  & 13.36 & 329.0 & 1.88 & \\
\midrule
Intel i7 11th gen & CPU (Iris Xe) & F16
  & 69.28 & est.\ 1{,}385\,J & est.\ 0.088
  & \multirow{2}{*}{${\approx}6.4\times$ (Q4)} \\
Intel i7 11th gen & CPU (Iris Xe) & Q4\_K\_M
  & 27.06 & est.\ 541\,J & est.\ 0.561 & \\
\midrule
\multicolumn{7}{l}{\footnotesize Laptop energy estimated at 20\,W
  sustained package power; LpW ratio is power-invariant
  (see Section~D.4).} \\
\multicolumn{7}{l}{\footnotesize T4 values from primary study
  ($n = 500$). Laptop values from this appendix ($n = 100$).} \\
\bottomrule
\end{tabular}
\end{table}

\subsection*{D.5 Pedagogical Latency Implications}

From the perspective of cognitive load and learner flow, the laptop
results present a qualitatively different picture from the T4. Under
F16, a mean latency of 69.3\,s is well beyond any pedagogically
tenable threshold: flow theory and CLT research consistently identify
latencies above 10--15\,s as disruptive to working memory continuity
\cite{Csikszentmihalyi1990, Sweller2011}, and 70\,s responses would
effectively preclude real-time tutoring use. Under Q4\_K\_M, a mean
latency of 27.1\,s remains above the preferred threshold but
represents a qualitatively different deployment scenario: a student
could plausibly interact with a Q4\_K\_M model on a consumer laptop
in a self-paced, asynchronous context, whereas F16 on the same
hardware is untenable for any interactive use case. This stands in
contrast to the T4 result, where both FP16 (9.2\,s) and NF4
(13.4\,s) sit within or near the pedagogically tenable window and
the deployment decision turns on a 4.2\,s latency difference rather
than a 42.2\,s one.

\subsection*{D.6 Implications for the Hardware-Dependence Argument}

The cross-platform comparison yields a finding that directly extends
the primary study's inference-regime dependence result: not only does
the FP16--NF4 efficiency relationship vary with KV-cache
configuration on a single platform ($1.33\times$ vs.\ $7.4\times$
on the T4), it also \emph{reverses sign entirely} across hardware
architectures. On a Turing GPU where dequantisation overhead is the
binding constraint, FP16 achieves higher LpW. On a consumer CPU
where memory bandwidth is the binding constraint, quantisation
achieves substantially higher LpW (${\approx}6.4\times$). Both
outcomes are predicted by the same underlying framework: the
efficiency of low-bit quantisation depends on whether the target
hardware can exploit weight compression directly (CPU, memory-bound)
or must pay a precision-restoration penalty at each decoding step
(Turing GPU, compute-bound without INT4 tensor cores).

The practical implication for educational AI deployment is direct:
efficiency claims derived from one hardware platform---whether a
cloud GPU benchmark or a consumer CPU test---cannot be extrapolated
to other platforms without empirical validation. Schools and
policymakers procuring AI tutoring hardware should insist on
efficiency measurements taken on the specific devices learners will
use, under realistic inference conditions, rather than relying on
vendor benchmarks or research results from mismatched hardware. For
the low-resource settings the paper's equity argument most directly
addresses, where consumer laptops and low-cost devices are the
realistic deployment target, the present results suggest that
quantised models under \texttt{llama.cpp} or equivalent CPU-optimised
runtimes are likely to outperform full-precision alternatives on
both LpW and pedagogical latency grounds---precisely the opposite
conclusion from the T4-only results.

\end{appendices}

\end{document}